\documentclass[aps, prd, twocolumn, superscriptaddress, nolongbibliography, nofootinbib]{revtex4-2}

\usepackage{xcolor}
\usepackage{amsmath}
\usepackage{siunitx}
\usepackage{graphicx}

\colorlet{darkBlue}{blue!50!black}

\usepackage[colorlinks=true, pdfborder={0 0 0}, linkcolor=darkBlue, citecolor=darkBlue, urlcolor=black, bookmarks, pdftitle={}]{hyperref}

\newcommand{\hairsp}{\hspace{1pt}}

\defcitealias{Saurenhaus2026}{Paper I}

\begin{document}

\title{Detectability of Seyfert galaxies in point-source and stacking searches with a global network of neutrino telescopes}

% author list
\author{Lena Saurenhaus}
\email[]{lsaurenh@mpp.mpg.de}
\affiliation{Max Planck Institute for Physics, 85748 Garching, Germany}
\affiliation{Department of Physics, TUM School of Natural Sciences, Technical University of Munich, 85747 Garching, Germany}

\author{Francesca Capel}
\affiliation{Max Planck Institute for Physics, 85748 Garching, Germany}

\author{Foteini Oikonomou}
\affiliation{Institutt for fysikk, Norwegian University of Science and Technology, 7491 Trondheim, Norway}

\date{\today}

% -------------------------------------------------------------------------------------------------

\begin{abstract}

Recent evidence for high-energy neutrino emission from NGC~1068 and other nearby Seyfert galaxies suggests that active galactic nuclei (AGNs) may constitute a class of astrophysical neutrino sources. Using a physically motivated neutrino emission model based on stochastic particle acceleration in magnetized AGN coronae, we investigate the detection prospects of X-ray-bright Seyfert galaxies with a detector network comprising IceCube, Baikal-GVD, KM3NeT/ARCA, and P-ONE. We find that the Circinus Galaxy is the most promising point-source target in the Southern sky and will be detected at the $5\sigma$ level by 2038 at the latest, provided that the Seyfert population accounts for the diffuse astrophysical neutrino flux up to ${\sim}\hairsp 100\,\mathrm{TeV}$ and Circinus is at least as efficient a neutrino emitter as the average source in the population. Individual detections of other nearby Seyferts are expected to remain challenging, making stacking searches informed by X-ray observations the most promising approach for discovering them within the next decade. Our results demonstrate that joint analyses across multiple neutrino telescopes, together with future MeV $\gamma$-ray observations, will be essential for establishing Seyfert galaxies as astrophysical neutrino sources and provide a means for testing the coronal origin of their neutrino emission.

\end{abstract}

\maketitle

% -------------------------------------------------------------------------------------------------

\section{Introduction}

The evidence for high-energy neutrino emission from the nearby Seyfert galaxy NGC~1068 presented by the IceCube Collaboration \cite{Abbasi2022} has opened a promising new direction in the search for the sources of astrophysical neutrinos. Subsequent studies have highlighted Seyfert galaxies as a potentially significant source class in the TeV energy range, and recent stacking analyses in both the Northern and Southern skies have further explored their collective contribution \cite{Abbasi2025b,Abbasi2026}. The observations of NGC~1068 and the absence of accompanying TeV $\gamma$-ray emission, despite its neutrino signal, suggest that the neutrinos originate from a ``hidden'' environment where high-energy photons are efficiently absorbed. This behavior is naturally accommodated in models in which particle acceleration and hadronic interactions occur in the corona of the active galactic nucleus (AGN) close to the supermassive black hole (SMBH) \cite{Padovani2024a,Murase2020}. If confirmed as a neutrino-emitting source class, Seyfert galaxies would provide a unique probe of particle acceleration and radiation processes in the immediate vicinity of actively accreting black holes, with important implications for multi-messenger astrophysics.

The potential contribution of Seyfert galaxies to the diffuse astrophysical neutrino flux has been investigated in several different studies, using a variety of modeling assumptions and generally finding results that are broadly consistent with the latest IceCube observations \cite{Padovani2024b,Ambrosone2024,Fiorillo2025,Karavola2026,Murase2026,Yang2026,Testagrossa2026,Carpio2026}. In our previous work \cite{Saurenhaus2026} (hereafter \citetalias{Saurenhaus2026}), we developed a physically motivated population model for neutrino emission from Seyfert galaxies based on stochastic proton acceleration in magnetized AGN coronae, in which the neutrino luminosity is tied to the intrinsic X-ray emission of each source. Our approach consistently combined a realistic coronal emission model with population synthesis and multi-messenger constraints. By fitting the public IceCube data for NGC~1068 and comparing the associated cascaded $\gamma$-ray emission to Fermi-LAT observations, we found that the neutrino-emitting region must be compact, with a characteristic coronal radius of $R_c \,{\sim}\, 5\,R_\mathrm{S}$. Similarly compact emission regions have since been inferred by \citet{Carpio2026} for several nearby Seyfert galaxies. Extending our model to the full Seyfert population, we further showed that NGC~1068 cannot be representative of the broader source population. If all Seyferts shared the best-fit parameter values inferred for NGC~1068, their cumulative neutrino emission would exceed current constraints on the diffuse flux at TeV energies by $3.8\sigma$. These results imply that those Seyfert galaxies that start to emerge as neutrino point sources must be exceptionally powerful neutrino emitters within a broader, less efficient population. Other works have come to similar conclusions, requiring a distribution of source parameters across the population \cite{Karavola2026}, suppression factors \cite{Yang2026,Testagrossa2026}, or an X-ray luminosity function of neutrino-emitting AGNs that deviates from that of the full AGN population \cite{Carpio2026} in order to remain consistent with the observed diffuse neutrino flux.

These findings motivate a natural next step. If Seyfert galaxies are not uniformly efficient neutrino emitters, but instead comprise a population in which only a small fraction of sources resemble NGC~1068, what are the prospects for detecting them as a class? In this work, we address this question by using the physically constrained population model of \citetalias{Saurenhaus2026} to forecast the detectability of Seyfert galaxies in both point-source and stacking searches, building on previous work that has highlighted the potential of stacking analyses to accelerate the discovery of populations of neutrino sources \cite[e.g.,][]{Kheirandish2021,Kowalski2025,Bartos2025}. Rather than asking whether Seyfert galaxies  can explain the observed diffuse flux, we investigate under what conditions individual sources and their collective emission can be identified with current and next-generation neutrino telescopes. We further explore the potential gains from combining observatories with complementary effective areas and sky coverage, as envisioned in future coordinated analyses \cite[e.g.,][]{Schumacher2025}. Our approach places the neutrino emission scenario for Seyfert galaxies in a predictive framework, linking population-level constraints directly to concrete observational strategies. 

We start in Section~\ref{sec:nu_emission_scenarios} by briefly reviewing the population model developed in \citetalias{Saurenhaus2026} and defining the representative population scenarios considered here, together with their implications for the diffuse neutrino emission from the Seyfert population. In Section~\ref{sec:detectors}, we introduce the present and future neutrino telescopes considered in this work and explain how we model their properties, while in Section~\ref{sec:point_src}, we explore the detectability of individual nearby Seyfert galaxies and discuss the role of MeV $\gamma$-ray observations as an independent probe of the coronal emission region. Section~\ref{sec:stacking} presents forecasts for stacking analyses, quantifies the dependence of the stacking significance on the source selection, and outlines an optimized analysis strategy. Finally, we summarize our results and discuss their implications in Section~\ref{sec:conclusions}.

% -------------------------------------------------------------------------------------------------

\section{Population model and neutrino emission scenarios}
\label{sec:nu_emission_scenarios}

Our neutrino emission model for Seyfert galaxies presented in \citetalias{Saurenhaus2026} is based on the assumption that inside the AGN corona, protons undergo stochastic particle acceleration in magnetized turbulence and subsequently produce high-energy neutrinos and $\gamma$-rays in photohadronic interactions with coronal X-rays and hadronuclear interactions with ambient gas. While neutrinos can escape the corona unhindered, $\gamma$-rays are attenuated in $\gamma\gamma$ interactions with X-rays and are thus expected to be observable only in the MeV range. Besides the intrinsic $2\,\text{--}\,10\,\mathrm{keV}$ X-ray luminosity, $L_X$, the neutrino and $\gamma$-ray emission of a source depends on the ratio of the cosmic-ray pressure to the thermal gas pressure within the corona, $P_\mathrm{CR}/P_\mathrm{th}$, the inverse turbulence strength, $\eta$, which is defined as
\begin{equation}
    \eta = \frac{B^2}{\delta B^2} = \frac{B_0^2 + \delta B^2}{\delta B^2} \,,
\end{equation}
and the coronal radius $R_c = r_c R_S$. Here, $B_0$ and $\delta B$ denote the regular and the turbulent component of the coronal magnetic field, respectively, and $R_S = 2GM_\mathrm{BH}/c^2$ is the Schwarzschild radius of the central SMBH.

In \citetalias{Saurenhaus2026}, we applied our model to a source population simulated based on the X-ray luminosity function of AGNs by \citet{Ueda2014} in order to calculate the expected diffuse neutrino emission from the Seyfert population. Comparing our predictions to the latest IceCube measurements of the diffuse astrophysical neutrino flux, we derived constraints on the free model parameters $P_\mathrm{CR}/P_\mathrm{th}$ and $\eta$. Our results indicate that it is highly unlikely that all sources are described by the same parameter values. Instead, they suggest that those Seyfert galaxies that emerge as point sources in the IceCube data are unusually powerful neutrino emitters with exceptional coronal properties, while the majority of sources must be considerably less efficient in order for the resulting diffuse neutrino flux not to exceed the IceCube observations.

Since these results disfavor a single set of parameter values describing the entire population, we explore three different neutrino emission scenarios to assess how the detectability of nearby, X-ray-bright Seyfert galaxies depends on their assumed coronal properties:
\begin{enumerate}
    \item The \textit{NGC~1068-like scenario}, in which all Seyfert galaxies are assumed to be described by the best-fit parameter values obtained for NGC~1068 in \citetalias{Saurenhaus2026}, $P_\mathrm{CR}/P_\mathrm{th} = 0.5$ and $\eta = 56$.
    \item The \textit{average scenario}, in which all Seyfert galaxies have a pressure ratio of $P_\mathrm{CR}/P_\mathrm{th} = 0.01$ and an inverse turbulence strength of $\eta = 1$. 
    \item The \textit{optimistic scenario}, in which all Seyfert galaxies are assumed to emit neutrinos at the highest possible level allowed by our model. This corresponds to a pressure ratio of $P_\mathrm{CR}/P_\mathrm{th} = 0.5$ and an inverse turbulence strength of $\eta = 1$.
\end{enumerate}

\begin{figure}[b]
    \includegraphics[width=8.5cm]{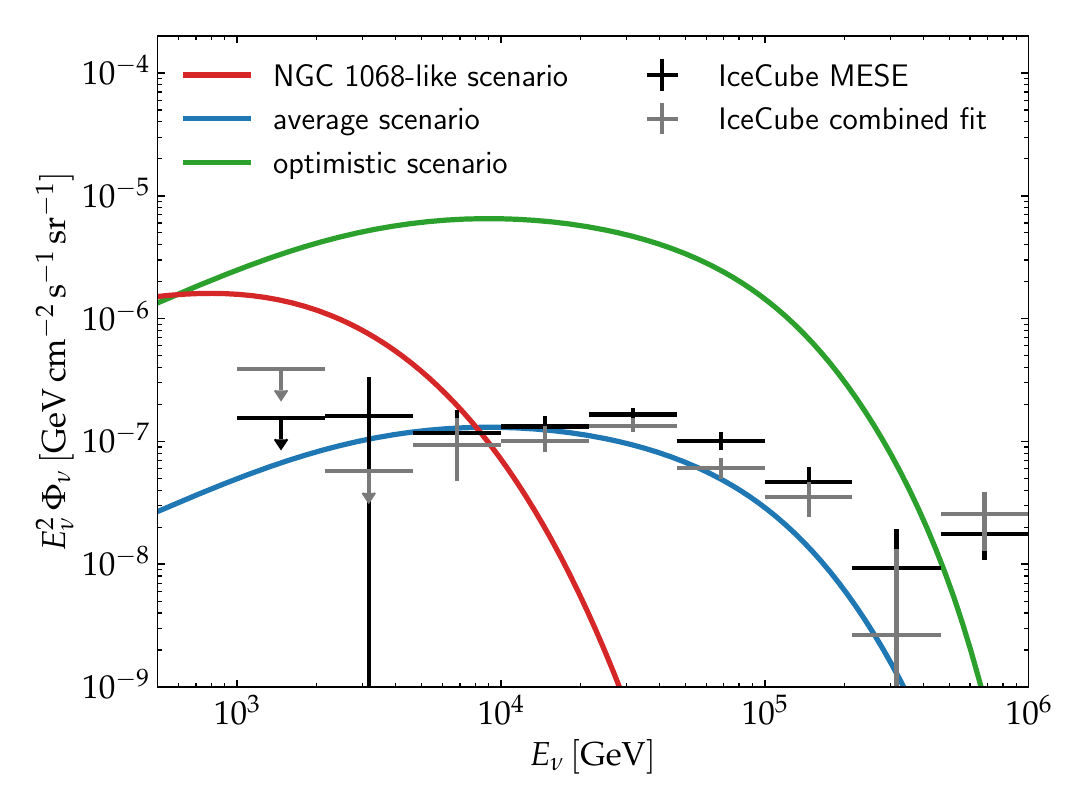}
    \caption{Diffuse all-flavor neutrino flux from the Seyfert population for the NGC~1068-like scenario (red), the average scenario (blue), and the optimistic scenario (green). The diffuse astrophysical neutrino flux measured by IceCube in the MESE analysis and the combined fit \cite{Abbasi2025a} is shown for comparison.}
    \label{fig:diff_fluxes}
\end{figure}

Figure~\ref{fig:diff_fluxes} shows the spectra of the diffuse neutrino flux from the Seyfert population for these three scenarios together with the latest measurements of the diffuse astrophysical neutrino flux by IceCube \cite{Abbasi2025a}. Notably, in the average scenario, our model predictions provide a good description of the IceCube observations up to energies of ${\sim}\hairsp 100\,\mathrm{TeV}$. In contrast, in both the NGC~1068-like scenario and the optimistic scenario, the cumulative neutrino flux produced by the Seyfert population significantly exceeds the IceCube observations. Motivated by the results of our model fits for NGC~1068, NGC~4151, and CGCG~420-015 reported in \citetalias{Saurenhaus2026}, which have shown that a small number of sources in the population may be above-average neutrino emitters, we consider these two scenarios nonetheless.

% -------------------------------------------------------------------------------------------------

\section{Neutrino telescope network}
\label{sec:detectors}

As suggested in Ref.~\cite{Schumacher2025}, we consider a network of neutrino telescopes with complementary sky coverage, comprising IceCube, Baikal-GVD, KM3NeT/ARCA, and P-ONE. While IceCube is located at the South Pole and has been operating in its full configuration since 2011, both Baikal-GVD and KM3NeT/ARCA are currently under construction in the Northern hemisphere and have already started taking data in their partial detector configurations. Baikal-GVD is located in Lake Baikal in Russia and currently consists of 117 strings covering a detector volume of $0.7\,\mathrm{km}^3$ \cite{Safronov2025}. It is planned to have an instrumented volume of $1\,\mathrm{km}^3$, which is expected to be achieved in 2028. KM3NeT/ARCA is under construction in the Mediterranean off the coast of Sicily \cite{AdrianMartinez2016}. As of July 2025, 51 vertical detection lines have been deployed, and the detector is anticipated to start operating in its full configuration comprising 230 detection lines covering a volume of $1\,\mathrm{km}^3$ in 2030 \cite{Dornic2025}. In addition, P-ONE is planned to be constructed in the Pacific Ocean in the Cascadia Basin off the coast of Canada \cite{Agostini2020}. Two pathfinder experiments, STRAW \cite{Boehmer2019} and STRAW-b \cite{Holzapfel2024}, with the aim of studying the optical properties of the sea water at the future detector site have already been completed successfully. P-ONE-1, the first instrumented line of the detector, is under construction and planned to be deployed this year \cite{Aghaei2026}. For the following calculations, we assume that P-ONE will start operating in its full size of $1\,\mathrm{km}^3$ in 2036, i.e.~in ten years from now. The properties and anticipated start dates of all neutrino telescopes considered in this work are summarized in Table~\ref{tab:detector_props}.

\begin{table}
    \caption{Properties and anticipated start dates of the neutrino telescopes considered in this work. The coordinates of the different detectors are taken from Ref.~\cite{Schumacher2025}.}
    \begin{ruledtabular}
    \begin{tabular}{lccc}
        & coordinates & size & start date \\
        \hline
        \noalign{\vskip 1ex}
        IceCube & $0^\circ\,\mathrm{E}$, $-90^\circ\,\mathrm{N}$ & $1\,\mathrm{km}^3$ & 2011 \\
        Baikal-GVD & $108.17^\circ\,\mathrm{E}$, $53.56^\circ\,\mathrm{N}$ & $1\,\mathrm{km}^3$ & 2028 \\
        KM3NeT/ARCA & $16.6^\circ\,\mathrm{E}$, $36.27^\circ\,\mathrm{N}$ & $1\,\mathrm{km}^3$ & 2030 \\
        P-ONE & $-127.73^\circ\,\mathrm{E}$, $47.74^\circ\,\mathrm{N}$ & $1\,\mathrm{km}^3$ & 2036 \\
    \end{tabular}
    \end{ruledtabular}
    \label{tab:detector_props}
\end{table}

Since only limited detector information is publicly available for Baikal-GVD, KM3NeT/ARCA, and P-ONE, we use the joint analysis framework \texttt{PLE$\nu$M} \cite{Schumacher2025} to model these three detectors based on the instrument response function of IceCube that is provided as part of the public 10-year dataset of track-like events for point-source searches \cite{IceCube2021}. Baikal-GVD, KM3NeT/ARCA, and P-ONE are all planned to encompass an instrumented volume of $1\,\mathrm{km}^3$ in their final configurations. Thus, we assume that all of them have the same effective area as IceCube, simply rotated to their specific location on Earth given in Table~\ref{tab:detector_props}. In addition, each detector is described by an energy resolution function, which provides a mapping between the true neutrino energy and the reconstructed muon energy, and an angular resolution function, which provides a measure for how close a muon is reconstructed to the arrival direction of its parent neutrino. For all four detectors considered in this work, we use the baseline energy resolution function implemented in \texttt{PLE$\nu$M}. It is calculated from the detector response matrix published as part of the 10-year IceCube data release by taking the average over the horizontal ($-10^\circ \leq \delta \leq 10^\circ$) and the upgoing ($10^\circ \leq \delta \leq 90^\circ$) declination band. 

By default, the \texttt{PLE$\nu$M} framework also suggests using the publicly available angular resolution function of IceCube for all four detectors. However, while IceCube is built in the Antarctic ice, Baikal-GVD, KM3NeT/ARCA, and P-ONE use water as their detection medium. Due to the longer scattering length of Cherenkov photons in water, these detectors should achieve a better angular resolution than IceCube, especially at neutrino energies $E_\nu \,{\gtrsim}\,10\,\mathrm{TeV}$. To take this into account, we model the angular resolution function of KM3NeT/ARCA separately using the angular deviation reported in Ref.~\cite{Aiello2024}. This is explained in more detail in Appendix~\ref{sec:ang_res_KM3NeT}. Although the water composition at Baikal-GVD and P-ONE is expected to differ from that at KM3NeT/ARCA, individual measurements suggest that the attenuation length of Cherenkov photons is comparable at all three detector sites \cite{Avrorin2019,Bailly2021,Riccobene2007}. Thus, we assume for our following calculations that Baikal-GVD and P-ONE have the same angular resolution function as KM3NeT/ARCA.

\begin{table*}
    \caption{Properties of the ten Seyfert galaxies with the highest intrinsic $2\,\text{--}\,10\,\mathrm{keV}$ X-ray flux in the BASS catalog \cite{Ricci2017} (excl.~NGC~1068). The luminosity distances are calculated from the redshifts given in the catalog. Only for sources with $d_L \leq 10\,\mathrm{Mpc}$, we use distances from redshift-independent measurements and recalculate the intrinsic X-ray luminosity accordingly. Here, $L_X$ is given in units of erg/s.}
    \begin{ruledtabular}
    \begin{tabular}{lSScSSS}
        & \multicolumn{1}{c}{$\mathrm{RA}\,[^\circ]$} & \multicolumn{1}{c}{$\delta\, [^\circ]$} & \multicolumn{1}{c}{z} & \multicolumn{1}{c}{$d_L\, [\mathrm{Mpc}]$} & {$F_X\, [10^{-10}\,\mathrm{erg}\,\mathrm{cm}^{-2}\,\mathrm{s}^{-1}]$} & {$\log_{10} L_X$} \\
        \hline
        \noalign{\vskip 1ex}
        Circinus Galaxy & 213.29 & -65.34 & 0.00145 & \multicolumn{1}{c}{\,\,\,4.2\footnotemark[1]} & 9.84 & 42.32 \\
        ESO~138-1 & 252.85 & -59.23 & 0.00914 & 39.4 & 6.71 & 44.09 \\
        NGC~7582 & 349.60 & -42.37 & 0.00488 & 21.0 & 5.08 & 43.48 \\
        NGC~424 & 17.87 & -38.08 & 0.01088 & 47.0 & 1.88 & 43.77 \\
        CGCG~164-019 & 221.40 & 27.03 & 0.02959 & 129.6 & 1.80 & 44.57 \\
        NGC~7212 & 331.76 & 10.23 & 0.02667 & 116.5 & 1.58 & 44.00 \\
        NGC~4945 & 196.36 & -49.47 & 0.00188 & \multicolumn{1}{c}{\,\,\,3.7\footnotemark[2]} & 1.49 & 41.39 \\
        NGC~3783 & 174.76 & -37.74 & 0.00965 & 41.6 & 1.29 & 43.43 \\
        IC~4329A & 207.33 & -30.31 & 0.01591 & 67.0 & 1.19 & 43.83 \\
        NGC~1194 & 45.95 & -1.10 & 0.01357 & 58.7 & 1.18 & 43.69 \\
    \end{tabular}
    \end{ruledtabular}
    \footnotetext[1]{From \cite{Karachentsev2013}.}
    \footnotetext[2]{From \cite{Tully2013}.}
    \label{tab:src_properties}
\end{table*}

In addition to IceCube, Baikal-GVD, KM3NeT/ ARCA, and P-ONE, several larger neutrino telescopes have been proposed for construction in the South China Sea. These include TRIDENT \cite{Ye2023} and NEON \cite{Zhang2025} with planned instrumented detector volumes of $7.5\,\mathrm{km}^3$ and $10\,\mathrm{km}^3$, respectively, as well as HUNT \cite{Chen2026}, which is projected to reach an instrumented volume of $30\,\mathrm{km}^3$\footnote{For HUNT, Lake Baikal has been proposed as an alternative location.}. While R\&D and prototype activities are underway, in particular for TRIDENT, the final configurations and construction timelines of these detectors remain uncertain. Therefore, we do not include them in our analysis.

% -------------------------------------------------------------------------------------------------

\section{Detectability of Seyfert galaxies as individual point sources}
\label{sec:point_src}

Using the \texttt{PLE$\nu$M} framework with the modifications to the detector modeling outlined in Section~\ref{sec:detectors}, we evaluate the detectability of a selection of X-ray-bright Seyfert galaxies with a neutrino telescope network comprising IceCube, Baikal-GVD, KM3NeT/ARCA, and P-ONE, considering the three different neutrino emission scenarios introduced in Section~\ref{sec:nu_emission_scenarios}. In particular, we select the ten Seyfert galaxies from the BASS catalog \cite{Ricci2017} that exhibit the highest intrinsic X-ray flux in the $2\,\text{--}\,10\,\mathrm{keV}$  band and are not yet known to be sources of astrophysical neutrinos.\footnote{We only consider the ten brightest sources, since in a realistic scenario compatible with the diffuse neutrino flux observed by IceCube, all other sources are not expected to be detectable as individual point sources within a reasonable timeframe (see Section~\ref{sec:results_point_sources}).} Table~\ref{tab:src_properties} summarizes their properties, including their coordinates, distances, and intrinsic $2\,\text{--}\,10\,\mathrm{keV}$  X-ray fluxes and luminosities.

We do not consider NGC~6240, since this object is known to be the remnant of a galaxy merger containing multiple nuclei \cite{Kollatschny2020}, and its X-ray emission may therefore not originate from a typical AGN corona \cite{Paggi2022}. Moreover, we exclude Centaurus~A and NGC~1275, because they both feature powerful radio jets, and it is likely that their X-ray emission originates from these jets rather than from the AGN corona \cite{Rani2022,Rodi2023,Imazato2021}. Among the ten Seyfert galaxies listed in Table \ref{tab:src_properties}, the Circinus Galaxy is by far the source with the highest intrinsic X-ray flux. It does not have a particularly high intrinsic X-ray luminosity, but it is located very close to us at a distance of $4.2\,\mathrm{Mpc}$ \cite{Karachentsev2013}. Only NGC~1068, which is not listed here since we consider it a known neutrino source, exhibits an even higher intrinsic X-ray flux of $F_X = 2.8\times 10^{-9}\,\mathrm{erg}\,\mathrm{cm}^{-2}\,\mathrm{s}^{-1}$ in the $2\,\text{--}\,10\,\mathrm{keV}$ band.\footnote{This value of the intrinsic X-ray flux corresponds to $L_X = 4.17\times 10^{43}\,\mathrm{erg}/\mathrm{s}$ \cite{Marinucci2016} and $d_L = 11.14\,\mathrm{Mpc}$ \cite{Tikhonov2021}.} Of the ten Seyfert galaxies under investigation, only CGCG~164-019, NGC~7212, and NGC~1194 are located in the Northern sky ($\delta > -5^\circ$) and are thus detectable by IceCube in the throughgoing-muon-track channel. In contrast, due to their location in the Northern hemisphere far from the poles, Baikal-GVD, KM3NeT/ARCA, and P-ONE are sensitive to neutrino events from the directions of all ten sources.

\subsection{Constraints from $\gamma$-ray observations}

While no significant neutrino emission has been observed from any of the sources listed in Table \ref{tab:src_properties}, $\gamma$-ray observations (or non-observations) in the MeV range can provide indirect constraints on the allowed values of the model parameters $P_\mathrm{CR}/P_\mathrm{th}$ and $\eta$ and, consequently, the potential neutrino emission of a source. This is particularly relevant for the optimistic scenario, in which all sources are assumed to produce neutrinos and MeV $\gamma$-rays with the highest possible efficiency. Out of the ten sources considered here, only the Circinus Galaxy and NGC~4945 have been detected in $\gamma$-rays so far \cite{Murase2024}. 

\begin{figure}
    \includegraphics[width=8.5cm]{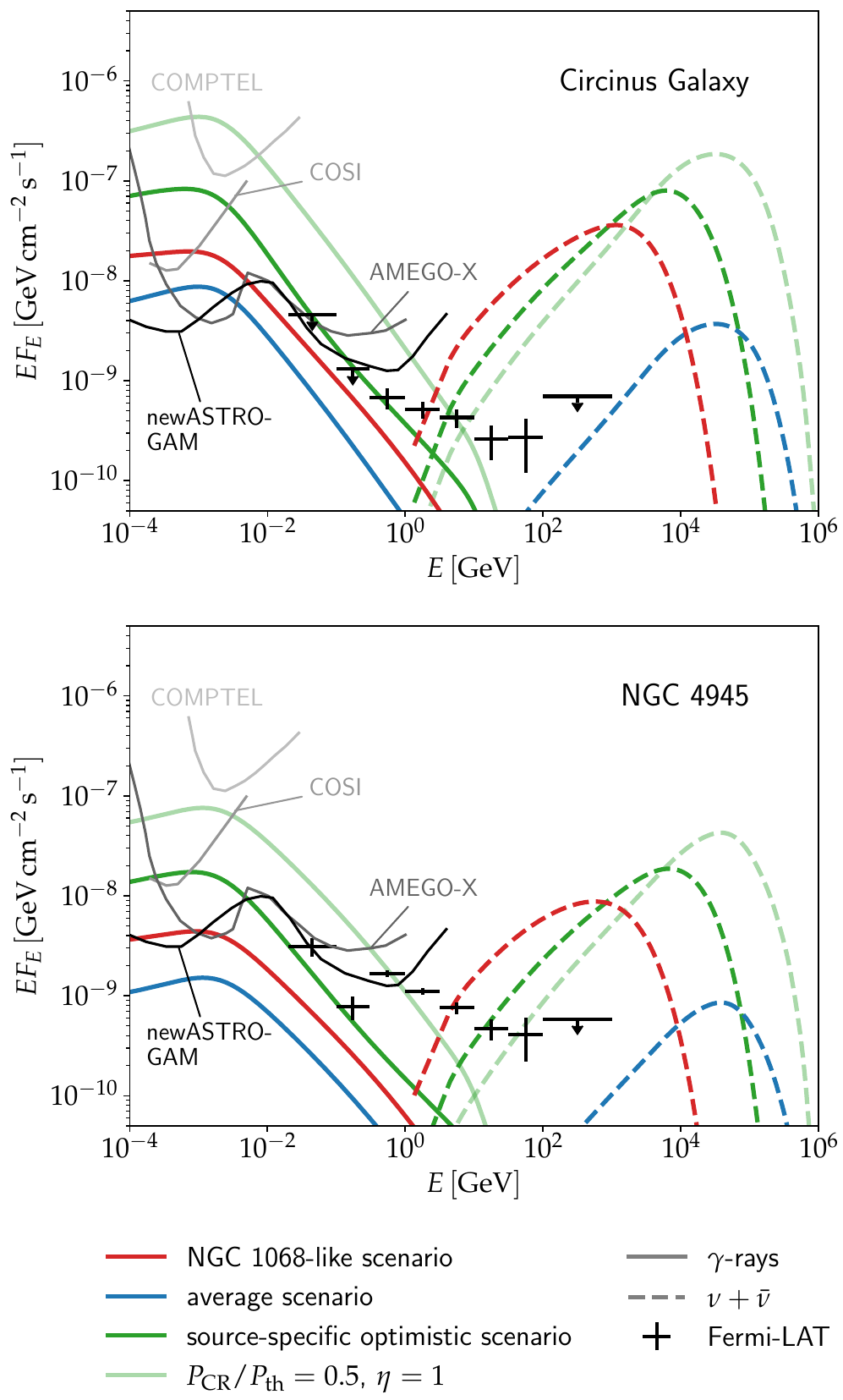}
    \caption{All-flavor neutrino spectra (dashed lines) and $\gamma$-ray spectra (solid lines) for the NGC~1068-like scenario (red), the average scenario (blue), and the source-specific optimistic scenario (dark green) for the Circinus Galaxy (top panel) and NGC~4945 (bottom panel). The light green dashed and solid lines correspond to the neutrino and $\gamma$-ray spectra for $P_\mathrm{CR}/P_\mathrm{th} = 0.5$ and $\eta = 1$. For each source, the $\gamma$-ray spectrum measured by Fermi-LAT \cite{Murase2024} is shown for comparison, with downward-pointing arrows indicating $95\hairsp\%$ upper limits. In addition, the sensitivity of COMPTEL \cite{Takahashi2013} is shown for its total lifetime of ${\sim}\hairsp 9\,\mathrm{years}$, while the sensitivity curves of COSI \cite{Tomsick2023}, AMEGO-X \cite{Caputo2022}, and newASTROGAM \cite{Berge2025} are shown for an observation time of $3\,\mathrm{years}$.}
    \label{fig:spectra_Circinus_NGC4945}
\end{figure}

Figure \ref{fig:spectra_Circinus_NGC4945} shows the neutrino spectra and the corresponding MeV $\gamma$-ray spectra of these two Seyfert galaxies for the three different scenarios together with their $\gamma$-ray spectra measured by Fermi-LAT and the sensitivity curves of past and future MeV $\gamma$-ray telescopes. For both sources, the $\gamma$-ray spectra calculated for $P_\mathrm{CR}/P_\mathrm{th} = 0.5$ and $\eta = 1$, which are shown as light green lines, are not compatible with the Fermi-LAT observations. In addition, neither of the two sources has been detected by the former MeV $\gamma$-ray telescope COMPTEL. Thus, for each of them, we define a source-specific optimistic scenario, distinct from the population-wide optimistic scenario introduced in Section~\ref{sec:nu_emission_scenarios}, for which our model predictions do not violate the Fermi-LAT observations and lie below the integrated sensitivity of COMPTEL \cite{Takahashi2013}. The resulting neutrino and MeV $\gamma$-ray spectra for this source-specific optimistic scenario are shown as dark green lines in Figure \ref{fig:spectra_Circinus_NGC4945} and correspond to $P_\mathrm{CR}/P_\mathrm{th} = 0.5$ and $\eta = 15$ for the Circinus Galaxy and $P_\mathrm{CR}/P_\mathrm{th} = 0.5$ and $\eta = 10$ for NGC~4945 (see also Table~\ref{tab:ps_detectability_opt_scen}). 

Alternatively, instead of increasing $\eta$, choosing a small\-er coronal radius could help to alleviate the tension between our model predictions and the Fermi-LAT observations (see \citetalias{Saurenhaus2026} for more details). However, the considered radius of $R_c = 5R_S$ is already quite small compared to that of the innermost stable orbit of a Schwarzschild black hole of $R_i = 3R_S$, which sets a lower limit on the size of the corona. Thus, we decide to keep $R_c$ fixed and adjust $\eta$ so that our model predictions agree with the $\gamma$-ray observations of the Circinus Galaxy and NGC~4945.

The remaining eight sources listed in Table~\ref{tab:src_properties} have not been detected by Fermi-LAT. However, for ESO~138-1 and NGC~7582, the non-detection of any MeV $\gamma$-ray emission by COMPTEL constrains the parameter values describing their source-specific optimistic scenario to $\eta = 2$ and $P_\mathrm{CR}/P_\mathrm{th} = 0.5$. For all other sources, our model predictions for the population-wide optimistic scenario ($\eta = 1$ and $P_\mathrm{CR}/P_\mathrm{th} = 0.5$) lie well below the integrated sensitivity of COMPTEL. The neutrino and MeV $\gamma$-ray spectra of these eight sources are shown in Figure~\ref{fig:spectra_8_sources} in Appendix~\ref{sec:nu_and_gamma_spec_remaining_srcs}, and the values of $\eta$ and $P_\mathrm{CR}/P_\mathrm{th}$ that describe their source-specific optimistic scenarios are summarized in Table~\ref{tab:ps_detectability_opt_scen}.

Figures~\ref{fig:spectra_Circinus_NGC4945} and \ref{fig:spectra_8_sources} also show the expected point-source sensitivities of COSI \cite{Tomsick2023}, a small $\gamma$-ray satellite scheduled to launch in 2027, and the proposed MeV $\gamma$-ray telescopes AMEGO-X \cite{Caputo2022} and newASTROGAM \cite{Berge2025}. If all Seyfert galaxies listed in Table~\ref{tab:src_properties} were correctly described by their source-specific optimistic scenario, COSI, AMEGO-X, and newASTROGAM would be able to detect the cascaded $\gamma$-ray emission from all of them within an observation time of 3 years. In contrast, under the assumption that all sources are described by the same parameter values as NGC~1068, only the $\gamma$-ray emission from the four brightest sources, the Circinus Galaxy, NGC~4945, ESO~138-1, and NGC~7582, would be detectable by AMEGO-X and newASTROGAM. The cascaded $\gamma$-ray flux from NGC~424 would lie right at the edge of the sensitivity curve of newASTROGAM, meaning that it could also be detected after an observation time of just over 3 years. With COSI having a significantly lower sensitivity than the other two telescopes, it would only be sensitive to the $\gamma$-ray emission from the Circinus Galaxy, the brightest of the ten Seyfert galaxies considered in this work. In the average scenario, Circinus would be the only source detectable by AMEGO-X and newASTROGAM, while all other Seyferts would remain undetected in the MeV range after 3 years of observation. These results highlight the importance of $\gamma$-ray observations in the MeV range for discriminating between different neutrino emission scenarios and for gaining a better understanding of neutrino production in Seyfert galaxies.

\subsection{Statistical method}
\label{sec:stat_method}

To estimate the detectability of a neutrino source with a given energy spectrum, we use the \texttt{PLE$\nu$M} framework \cite{Schumacher2025} and follow the statistical method implemented therein. We start by constructing an Asimov dataset \cite{Cowan2011} consisting of astrophysical neutrino events from the source and atmospheric background events, where the latter are calculated from the spectrum of the atmospheric neutrino flux computed with \texttt{MCEq} \cite{Fedynitch2015,Fedynitch2019}. This Asimov data is then binned in $\log\hat{E}$, where $\hat{E}$ denotes the reconstructed muon energy, and the squared angular distance between the reconstructed arrival direction of the muon and the position of the source under consideration, $\Delta\Psi^2 = |\hat{\Omega} - \Omega_\mathrm{src}|^2$. The total number of expected events in the $i$th reconstructed energy bin and the $j$th angular separation bin corresponds to the sum of the number of astrophysical and atmospheric events and is given by
\begin{equation}
    \lambda_{ij}(\boldsymbol{\theta}) = \lambda_{ij}^\mathrm{astro}(\boldsymbol{\theta}^\mathrm{astro}) + \lambda_{ij}^\mathrm{atm}(\phi_0^\mathrm{atm}) \,.
\end{equation}
Here, $\boldsymbol{\theta}^\mathrm{astro}$ denotes the free parameters of the spectral model, with $\boldsymbol{\theta}^\mathrm{astro} = (P_\mathrm{CR}/P_\mathrm{th}, \eta)$ for our model, and $\phi_0^\mathrm{atm}$ is the normalization of the atmospheric neutrino spectrum, which is also treated as a free parameter. Using \texttt{PLE$\nu$M}, we evaluate the agreement between our model predictions for parameter values $\boldsymbol{\theta} \equiv (\boldsymbol{\theta}^\mathrm{astro}, \phi_0^\mathrm{atm})$ and the Asimov data $\boldsymbol{x}$ by means of a binned Poisson likelihood of the form 
\begin{equation}
    \mathcal{L}(\boldsymbol{x}|\boldsymbol{\theta}) = \prod_{j=1}^{N_\Psi} \prod_{i=1}^{N_E} \frac{\left[ \lambda_{ij} (\boldsymbol{\theta}) \right]^{n_{ij}}}{n_{ij}!} e^{-\lambda_{ij} (\boldsymbol{\theta})} \,.
    \label{eq:L_single_detector}
\end{equation}
Here, $n_{ij}$ is the number of neutrino events in the $i$th energy bin and the $j$th angular separation bin in the Asimov dataset, and $N_E$ and $N_\Psi$ denote the total number of energy and angular separation bins, respectively. 

In order to extend this formalism to a network of multiple neutrino telescopes, the \texttt{PLE$\nu$M} framework treats detections of neutrino events originating from a single point source by different detectors as independent observations. Thus, the total likelihood function for multiple detectors can simply be written as the product of the likelihood functions for the individual detectors given by Eq.~(\ref{eq:L_single_detector}),
\begin{equation}
    \begin{aligned}
        \mathcal{L}_\mathrm{tot}(\boldsymbol{x}|\boldsymbol{\theta}) 
        & = \prod_{k=1}^{N_\mathrm{det}} \mathcal{L}_k (\boldsymbol{x}_k|\boldsymbol{\theta}) \\
        & = \prod_{k=1}^{N_\mathrm{det}} \prod_{j=1}^{N_\Psi} \prod_{i=1}^{N_E} \frac{\left[ \lambda_{ijk} (\boldsymbol{\theta}) \right]^{n_{ijk}}}{n_{ijk}!} e^{-\lambda_{ijk} (\boldsymbol{\theta})}\,.
    \end{aligned}
    \label{eq:single_source_lh}
\end{equation}
Here, $\lambda_{ijk}$ denotes the expected number of neutrino events detected by the $k$th detector in the $i$th reconstructed energy bin and the $j$th angular separation bin, $n_{ijk}$ is the respective number of neutrino events contained in the Asimov dataset $\boldsymbol{x}_k$ for the $k$th detector, and $k$ runs over all considered neutrino telescopes. To quantify the detectability of a source with a network of neutrino telescopes, we use the negative log-likelihood ratio as the test statistic,
\begin{equation}
    \mathrm{TS} = -2\log \frac{\mathcal{L}_\mathrm{tot}(\boldsymbol{x}| \hat{\phi}_0^\mathrm{atm}, P_\mathrm{CR}/P_\mathrm{th} = 0)}{\mathcal{L}_\mathrm{tot}(\boldsymbol{x}|\hat{\phi}_0^\mathrm{atm}, \hat{\boldsymbol{\theta}}^\mathrm{astro})} \,,
    \label{eq:single_source_TS}
\end{equation}
where $\mathcal{L}_\mathrm{tot}(\boldsymbol{x}| \hat{\phi}_0^\mathrm{atm}, P_\mathrm{CR}/P_\mathrm{th} = 0)$ is the likelihood maximized under the background hypothesis and $\mathcal{L}_\mathrm{tot}(\boldsymbol{x}|\hat{\phi}_0^\mathrm{atm}, \hat{\boldsymbol{\theta}}^\mathrm{astro})$ is the likelihood maximized under the signal hypothesis. Since our neutrino emission model has two free parameters, $P_\mathrm{CR}/P_\mathrm{th}$ and $\eta$, we use Wilks' theorem \cite{Wilks1938} with 2 degrees of freedom to convert the obtained TS value to a $p$-value. According to the convention of a two-sided Gaussian tail, we then transform this $p$-value into an equivalent significance in terms of Gaussian standard deviations.

\subsection{Results}
\label{sec:results_point_sources}

\begin{figure}
        \includegraphics[width=8.5cm]{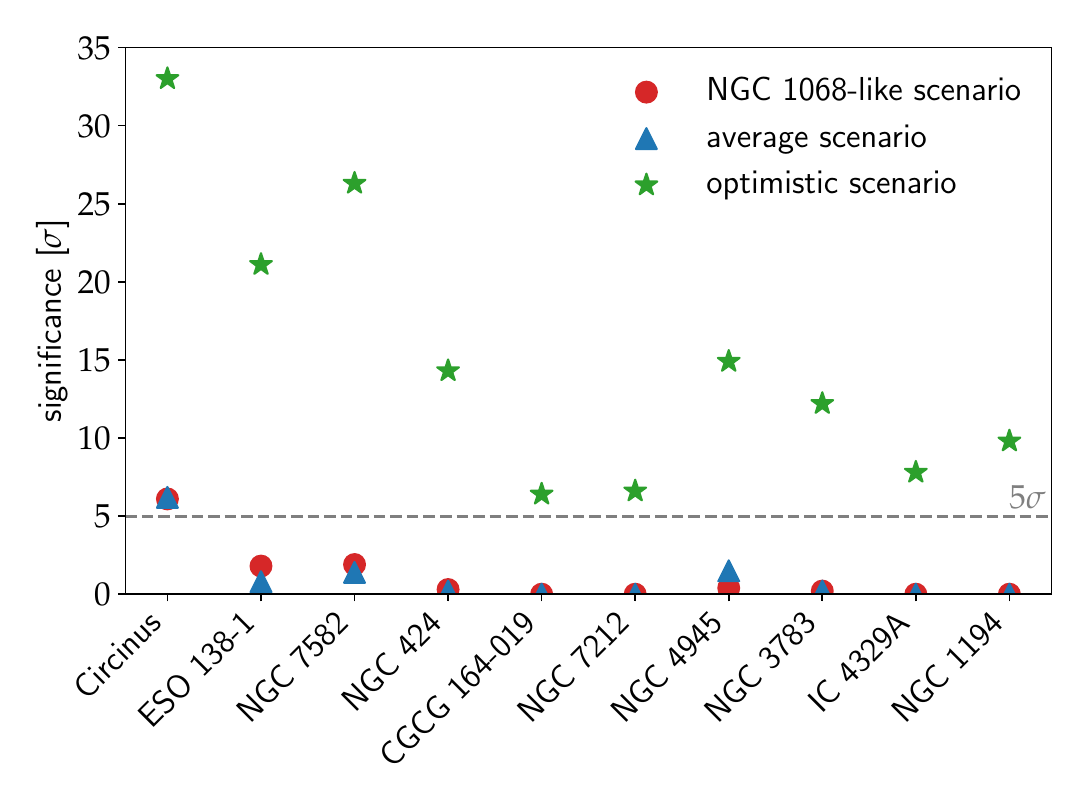}
        \caption{Significance at which each of the ten Seyfert galaxies listed in Table \ref{tab:src_properties} would be detected by a neutrino telescope network consisting of IceCube, Baikal-GVD, KM3NeT/ARCA, and P-ONE in 2040 in the NGC~1068-like scenario (red circles), the average scenario (blue triangles), and the most optimistic scenario compatible with the $\gamma$-ray observations of the respective source (green stars). The gray dashed line marks a significance of $5\sigma$.
        }
        \label{fig:detectability_10_Seyferts}
    \end{figure}

Using \texttt{PLE$\nu$M} and the statistical method outlined in Section \ref{sec:stat_method}, we investigate the detectability of the ten Seyfert galaxies listed in Table \ref{tab:src_properties} as individual point sources with a neutrino telescope network consisting of IceCube, Baikal-GVD, KM3NeT/ARCA, and P-ONE in the year 2040. Under the assumption that the latter three detectors start operating in their full configurations in 2028, 2030, and 2036, respectively, this corresponds to 29 years of IceCube, 12 years of Baikal-GVD, 10 years of KM3NeT/ARCA, and 4 years of P-ONE. Here, we do not take into account that Baikal-GVD and KM3NeT/ARCA are already recording data in their partial detector configurations. Our results for the three different neutrino emission scenarios introduced in Section \ref{sec:nu_emission_scenarios} are summarized in Figure \ref{fig:detectability_10_Seyferts}, with more details given in Tables \ref{tab:ps_detectability_NGC_like_and_avg_scen} and \ref{tab:ps_detectability_opt_scen} in Appendix \ref{sec:detailed_results}.

In all three considered scenarios, the Circinus Galaxy is expected to be by far the brightest Seyfert galaxy in the Southern sky, in both neutrinos and MeV $\gamma$-rays. In the most optimistic scenario compatible with the Fermi-LAT observations of the Circinus Galaxy, a significance of $5\sigma$ would already be reached in 2029, just one year after Baikal-GVD is anticipated to start operating in its full configuration. However, a recent point-source analysis utilizing 5 years of data recorded by Baikal-GVD in its current partial configuration, which would correspond to ${\sim}\hairsp 1.5\,\mathrm{years}$ of operation of the full detector, does not yield a significant detection \cite{Allakhverdyan2026}. This strongly suggests that the source-specific optimistic scenario does not provide an accurate description of the neutrino emission properties of the Circinus Galaxy.

\begin{figure*}
    \includegraphics[width=17.5cm]{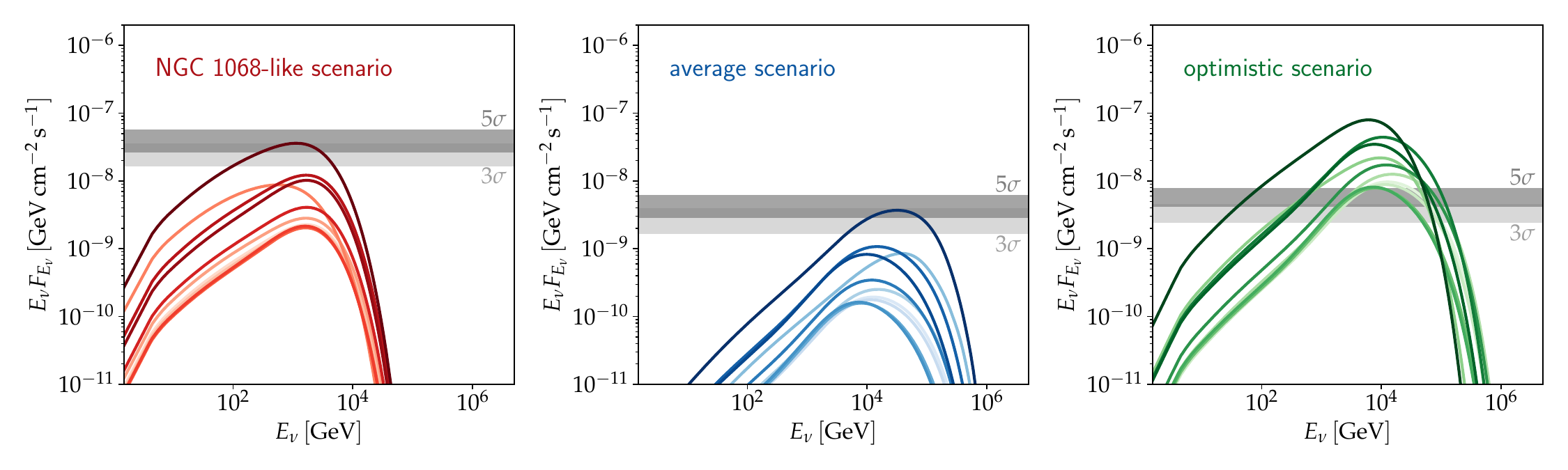}
    \caption{All-flavor neutrino spectra of the ten Seyfert galaxies listed in Table~\ref{tab:src_properties} for the NGC~1068-like scenario (left panel), the average scenario (middle panel), and the source-specific optimistic scenario (right panel). The spectra are colored according to the intrinsic $2\,\text{--}\,10\,\mathrm{keV}$ X-ray flux of the respective source, with darker colors corresponding to a higher flux. The light and dark gray bands indicate the flux ranges in which the peaks of the neutrino spectra would have to lie in order for the sources to be detected with a significance of $3\sigma$ and $5\sigma$ in 2040.}
    \label{fig:nu_spectra_brightest_Seyferts}
\end{figure*}

In both the NGC~1068-like scenario and the average scenario, the Circinus Galaxy would be detected with a significance of $6\sigma$ in 2040. More specifically, under the assumption that the Seyfert population is responsible for the diffuse astrophysical neutrino flux up to energies of ${\sim}\hairsp 100\,\mathrm{TeV}$ and the Circinus Galaxy is at least as bright as the average source in the population, it would be detected with a significance of $5\sigma$ latest in 2038. Thus, the Circinus Galaxy is one of the most promising targets for the next-generation neutrino telescopes currently under construction in the Northern hemisphere. This is in line with the results of a recent IceCube analysis searching for neutrino emission from Seyfert galaxies in the Southern sky \cite{Abbasi2026}. They find that, with a post-trial significance of $1.7\sigma$, the Circinus Galaxy is the most promising neutrino source candidate in their sample. In addition, they report a non-zero, but not statistically significant number of signal events associated with ESO~138-1 and NGC~7582, the second and the third brightest source on our list of Seyfert galaxies (see Table~\ref{tab:src_properties}). However, a direct comparison of these IceCube results to our predictions is difficult, mainly because the IceCube analysis is using starting-track events, while our detectability estimates are made for an analysis using throughgoing muon tracks. In this detection channel, the IceCube detector is not sensitive to any point source in the Southern sky due to the overwhelming background of atmospheric muons.

All other Seyfert galaxies on our list are predicted to be much fainter than the Circinus Galaxy, in both X-rays and neutrinos. In the NGC~1068-like scenario, a significance greater than $1\sigma$ would be achieved only for ESO~138-1 and NGC~7582, while in the average scenario this would only be the case for NGC~7582 and NGC~4945. For the remaining sources, it would take more than 35 years from now to reach a significance of $3\sigma$ in either of the two scenarios. Our detectability estimates for the average scenario differ only slightly from those for the NGC~1068-like scenario, although the underlying neutrino spectra are very different. As shown in Figure~\ref{fig:nu_spectra_brightest_Seyferts}, in the NGC~1068-like scenario the neutrino spectra have a much higher normalization than in the average scenario but peak at lower energies, where neutrino telescopes are less sensitive due to their smaller effective area and the more dominant atmospheric neutrino background. Thus, as indicated by the light- and dark-gray bands, in the NGC~1068-like scenario a much higher peak neutrino flux is required to reach a significance of $3\sigma$ or $5\sigma$ in 2040. This leads to similar results for both scenarios despite the overall lower number of signal events in the average scenario. 

In the source-specific optimistic scenario, all considered Seyfert galaxies would be detected at the $5\sigma$ level by 2035 (see Table~\ref{tab:ps_detectability_opt_scen}). However, given the results of the latest IceCube point-source analysis utilizing 13.1 years of data recorded between June 2010 and November 2023 \cite{Abbasi2025c}, it is rather unlikely that this scenario correctly describes the neutrino emission of CGCG~164-019, NGC~7212, and NGC~1194, all of which are located in the Northern sky. To facilitate a more accurate comparison between our model predictions and the IceCube observations of these three sources, we separately calculate the expected number of signal events as well as the significance that would be achieved after an observation time of 13.1 years\footnote{In our calculation, an observation time of 13.1 years corresponds to the year 2024, because for the sake of simplicity, we assume that the IceCube detector only started taking data in 2011 after it was completed.} with IceCube only. The results of this calculation are summarized in Table~\ref{tab:detectability_IC_only}. According to our predictions for the source-specific optimistic scenario, IceCube should have observed NGC~1194 as a neutrino point source at the $3\sigma$ level already in 2021. In addition, after 13.1 years of observations, a significance ${>}\hairsp 2.5\sigma$ should have been achieved for both CGCG~164-019 and NGC~7212. In contrast, based on a power-law fit, the IceCube Collaboration reports pre-trial significances of $1.8\sigma$ for NGC~1194, $0.9\sigma$ for CGCG~164-019, and $0.6\sigma$ for NGC~7212. Moreover, their analysis using the core-corona model by Kheirandish \textit{et al.}~\cite{Kheirandish2021} does not show evidence for neutrino emission from any of these three Seyfert galaxies. This suggests that the optimistic scenario overestimates the neutrino production efficiency of all three sources. Again, a direct comparison of our predictions to the IceCube results is not straightforward, since the latter are based on different spectral models and a different statistical method. In addition, part of the data used in the IceCube analysis was recorded before the completion of the full detector.

The latest point-source observations by IceCube and Baikal-GVD, which indicate that at least some of the Seyfert galaxies considered here must be much fainter in neutrinos than predicted in the source-specific optimistic scenario, are consistent with our previous finding that assuming this scenario for the entire Seyfert population would lead to a cumulative neutrino emission that significantly exceeds the observed diffuse astrophysical neutrino flux. Nevertheless, it makes sense to consider this scenario, since it is possible that a few, exceptional sources such as CGCG~420-015 are indeed correctly described by the underlying model assumptions, as shown in \citetalias{Saurenhaus2026}.

Neronov \textit{et al.}~\cite{Neronov2026} also study the detectability of a selection of nearby, X-ray-bright Seyfert galaxies with KM3NeT/ARCA. Using two different spectral templates based on observations of NGC~1068 and NGC~4151, they find that, in addition to these two sources, only the Circinus Galaxy and NGC~4945 are expected to be detectable by KM3NeT within 20 years of operation. This is roughly in agreement with the fact that in all three neutrino emission scenarios considered here, both the Circinus Galaxy and NGC~4945 are among the sources with the highest expected number of signal events (see Tables~\ref{tab:ps_detectability_NGC_like_and_avg_scen} and \ref{tab:ps_detectability_opt_scen}).

\begin{table}
    \caption{Detectability of the three Seyfert galaxies with the highest intrinsic $2\,\text{--}\,10\,\mathrm{keV}$ X-ray flux in the Northern sky (excl. NGC~1068) with IceCube alone after 13.1 years in the source-specific optimistic scenario ($\eta = 1$, $P_\mathrm{CR}/P_\mathrm{th} = 0.5$ for all three sources). We list the expected number of signal events, the TS value, the $p$-value together with the equivalent significance, and the years in which a significance of $3\sigma$ and $5\sigma$ would be reached.}
    \begin{ruledtabular}
    \begin{tabular}{lrrrcc}
        & \multicolumn{1}{c}{$n_s$} & \multicolumn{1}{c}{TS} & \multicolumn{1}{c}{$-\log_{10}p$} & $3\sigma$ in & $5\sigma$ in \\
        \hline
        \noalign{\vskip 1ex}
        CGCG~164-019 & 26.6 & 11.4 & 2.5 ($2.9\sigma$) & 2025 & 2045 \\
        NGC~7212 & 27.0 & 9.1 & 2.0 ($2.6\sigma$) & 2029 & 2053 \\
        NGC~1194 & 33.7 & 16.5 & 3.6 ($3.7\sigma$) & 2021 & 2034 \\
    \end{tabular}
    \end{ruledtabular}
    \label{tab:detectability_IC_only}
\end{table}

% -------------------------------------------------------------------------------------------------

\section{Prospects for stacking searches}
\label{sec:stacking}

In order for the population of Seyfert galaxies to account for the diffuse astrophysical neutrino flux up to energies of ${\sim}\hairsp 100\,\mathrm{TeV}$ without exceeding any IceCube measurements in the TeV range, the majority of sources in the population must have a low pressure ratio ($P_\mathrm{CR}/P_\mathrm{th} \lesssim 0.05$) and a high level of turbulence ($\eta \lesssim 10$) inside their coronae (see Appendix C of \citetalias{Saurenhaus2026}). The simplest scenario that meets these requirements is the average scenario introduced in Section~\ref{sec:nu_emission_scenarios}, in which all sources are assumed to have a pressure ratio of $P_\mathrm{CR}/P_\mathrm{th} = 0.01$ and an inverse turbulence strength of $\eta = 1$. However, in this scenario, only very few sources are predicted to be detectable within the next 35 years. Thus, the most promising method to search for the neutrino emission from Seyfert galaxies as a source class will likely be a stacking analysis.

\begin{table*}
    \caption{Detectability of the cumulative neutrino emission from all BASS catalog sources with $F_X \geq 10^{-11}\,\mathrm{erg}\,\mathrm{cm}^{-2}\,\mathrm{s}^{-1}$ including the Circinus Galaxy (top two rows; $N_\mathrm{src} = 213$) and excluding the Circinus Galaxy (bottom two rows; $N_\mathrm{src} = 212$) in a stacking analysis with a detector network consisting of IceCube, Baikal-GVD, KM3NeT/ARCA, and P-ONE in 2040. For both the NGC~1068-like scenario ($P_\mathrm{CR}/P_\mathrm{th} = 0.5$ and $\eta = 56$) and the average scenario ($P_\mathrm{CR}/P_\mathrm{th} = 0.01$ and $\eta = 1$), we list the expected total number of signal events, the TS value, the $p$-value together with the equivalent significance, and the years in which a significance of $3\sigma$ and $5\sigma$ would be reached.}
    \begin{ruledtabular}
    \begin{tabular}{lccccc}
        & $n_s$ & TS & $-\log_{10} p$ & $3\sigma$ in & $5\sigma$ in \\
        \hline
        \noalign{\vskip 1ex}
        NGC~1068-like scenario & 592.5 & 64.8 & 14.1 ($7.8\sigma$) & 2032 & 2035 \\
        average scenario & 139.5 & 60.8 & 13.2 ($7.5\sigma$) & 2032 & 2036 \\
        \noalign{\vskip 0.5ex}
        \hline
        \noalign{\vskip 1ex}
        NGC~1068-like scenario (excl. Circinus Galaxy) & 494.2 & 19.5 & 4.2 ($4.0\sigma$) & 2037 & 2045 \\
        average scenario (excl. Circinus Galaxy) & 113.5 & 17.5 & 3.8 ($3.8\sigma$) & 2038 & 2046 \\
    \end{tabular}
    \end{ruledtabular}
    \label{tab:results_stacking}
\end{table*}

We extend the statistical method presented in Section~\ref{sec:stat_method}, which we use to evaluate the detectability of individual Seyfert galaxies as neutrino point sources, to stacking analyses of multiple sources, assuming that observations of individual point sources are independent statistical processes. Under this assumption, we can simply multiply the values of the single-source likelihood defined in Eq.~(\ref{eq:single_source_lh}) for all sources included in the stacking analysis,
\begin{equation}
    \begin{aligned}
        \mathcal{L}_\mathrm{stack}(\boldsymbol{x}|\boldsymbol{\theta})
        & = \prod_{l=1}^{N_\mathrm{src}} \mathcal{L}_{\mathrm{tot},\hairsp l}(\boldsymbol{x}_l|\boldsymbol{\theta}) \\ 
        & = \prod_{l=1}^{N_\mathrm{src}} \prod_{k=1}^{N_\mathrm{det}} \prod_{j=1}^{N_\Psi} \prod_{i=1}^{N_E} \frac{\left[ \lambda_{ijkl} (\boldsymbol{\theta}) \right]^{n_{ijkl}}}{n_{ijkl}!} e^{-\lambda_{ijkl} (\boldsymbol{\theta})} \,,
    \end{aligned}
\end{equation}
where $\lambda_{ijkl}$ denotes the expected number of neutrino events from the $l$th source detected by the $k$th detector in the $i$th reconstructed energy bin and the $j$th angular separation bin, and $n_{ijkl}$ is the respective number of events in the Asimov dataset for the $l$th source, $\boldsymbol{x}_l$. Similar to the single-source case, we use the negative log-likelihood ratio,
\begin{equation}
    \mathrm{TS} = -2\log \frac{\mathcal{L}_\mathrm{stack}(\boldsymbol{x}|\hat{\phi}_0^\mathrm{atm}, P_\mathrm{CR}/P_\mathrm{th} = 0)}{\mathcal{L}_\mathrm{stack}(\boldsymbol{x}|\hat{\phi}_0^\mathrm{atm}, \hat{\boldsymbol{\theta}}^\mathrm{astro})}\,,
\end{equation}
as a test statistic to estimate the detectability of the neutrino emission from a sample of Seyfert galaxies in a stacking analysis. Again, we use Wilks' theorem with 2 degrees of freedom to convert the obtained TS value to a $p$-value, which we then convert to a significance in terms of Gaussian standard deviations according to the convention of a two-sided Gaussian tail.

For our stacking predictions, we select all Seyfert galaxies from the BASS catalog that have an intrinsic $2\,\text{--}\,10\,\mathrm{keV}$ X-ray flux greater than $F_X = 10^{-11}\,\mathrm{erg}\,\mathrm{cm}^{-2}\,\mathrm{s}^{-1}$, excluding NGC~1068, which we consider a known neutrino source. This results in a sample consisting of total of 213 sources, including those listed in Table~\ref{tab:src_properties}. Using the \texttt{PLE$\nu$M} framework, we investigate the detection prospects of the neutrino emission from this source sample in a stacking analysis combining data from IceCube, Baikal-GVD, KM3NeT/ARCA, and P-ONE. For that, we only consider the NGC~1068-like scenario and the average scenario and omit the optimistic scenario. In the latter, the cumulative neutrino emission from the Seyfert population would massively exceed the diffuse flux observations by IceCube (see Figure~\ref{fig:diff_fluxes}), which is why it is highly unlikely that more than just a few sources fulfill the assumptions made in this scenario.

Our predictions for the detectability of the neutrino emission from all Seyfert galaxies in our sample in a stacking analysis are summarized in Table~\ref{tab:results_stacking}. In the NGC~1068-like scenario, Seyfert galaxies would be discovered as a class of neutrino sources at the $5\sigma$ level in 2035, five years after the anticipated start date of KM3NeT/ARCA. In the average scenario, this would happen only one year later, in 2036. By 2040, a significance ${>}\hairsp 7\sigma$ would be reached in both scenarios. Although the detection prospects in the two scenarios are very similar, in the NGC~1068-like scenario a factor of ${\sim}\hairsp 4$ more signal events are required for the stacking analysis to yield an equally significant result as in the average scenario. As already discussed in Section~\ref{sec:results_point_sources}, this is due to the fact that in the NGC~1068-like scenario, Seyfert galaxies emit neutrinos at lower energies, where a neutrino telescope is less sensitive due to the higher background of atmospheric neutrinos and the reduced effective area.

\begin{figure}
\includegraphics[width=8.5cm]{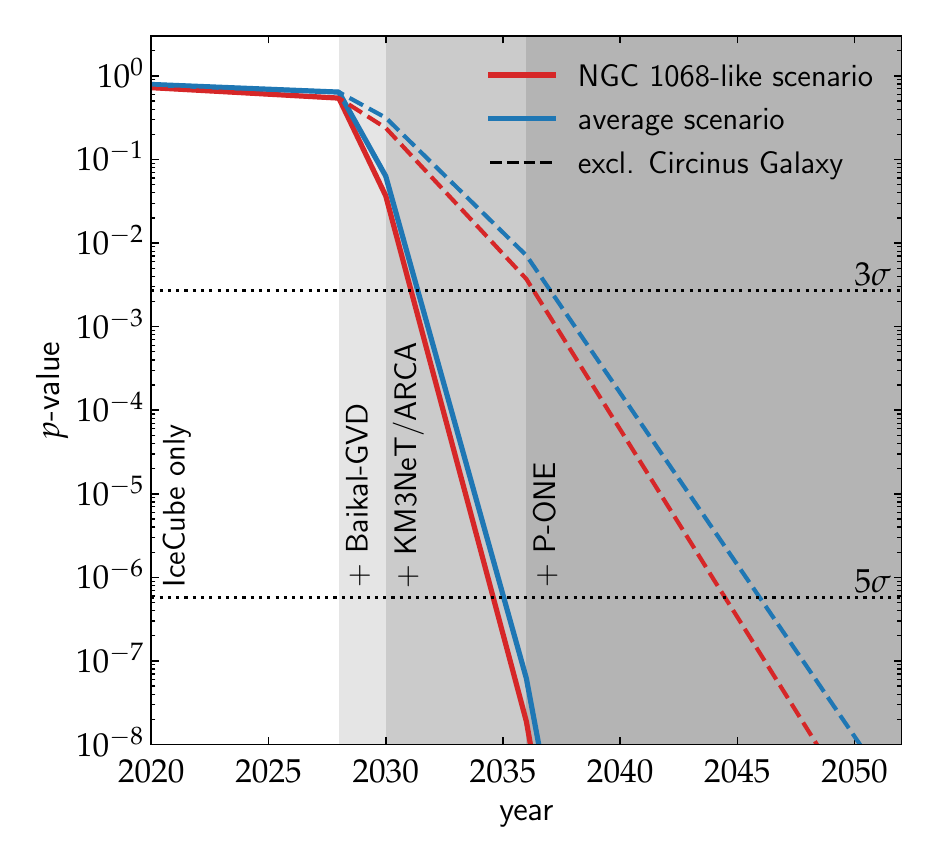}
\caption{$p$-value that would be achieved in a stacking analysis including all BASS catalog sources with $F_X \geq 10^{-11}\,\mathrm{erg}\,\mathrm{cm}^{-2}\,\mathrm{s}^{-1}$ in the NGC~1068-like scenario (red) and the average scenario (blue) including the Circinus Galaxy (solid lines) and excluding the Circinus Galaxy (dashed lines). The gray shaded areas indicate the start dates of those neutrino telescopes that are still under construction, and the black dotted lines mark the $p$-values that correspond to a significance of $3\sigma$ and $5\sigma$.}
\label{fig:p_val_vs_T}
\end{figure}

\begin{figure}
\includegraphics[width=8.5cm]{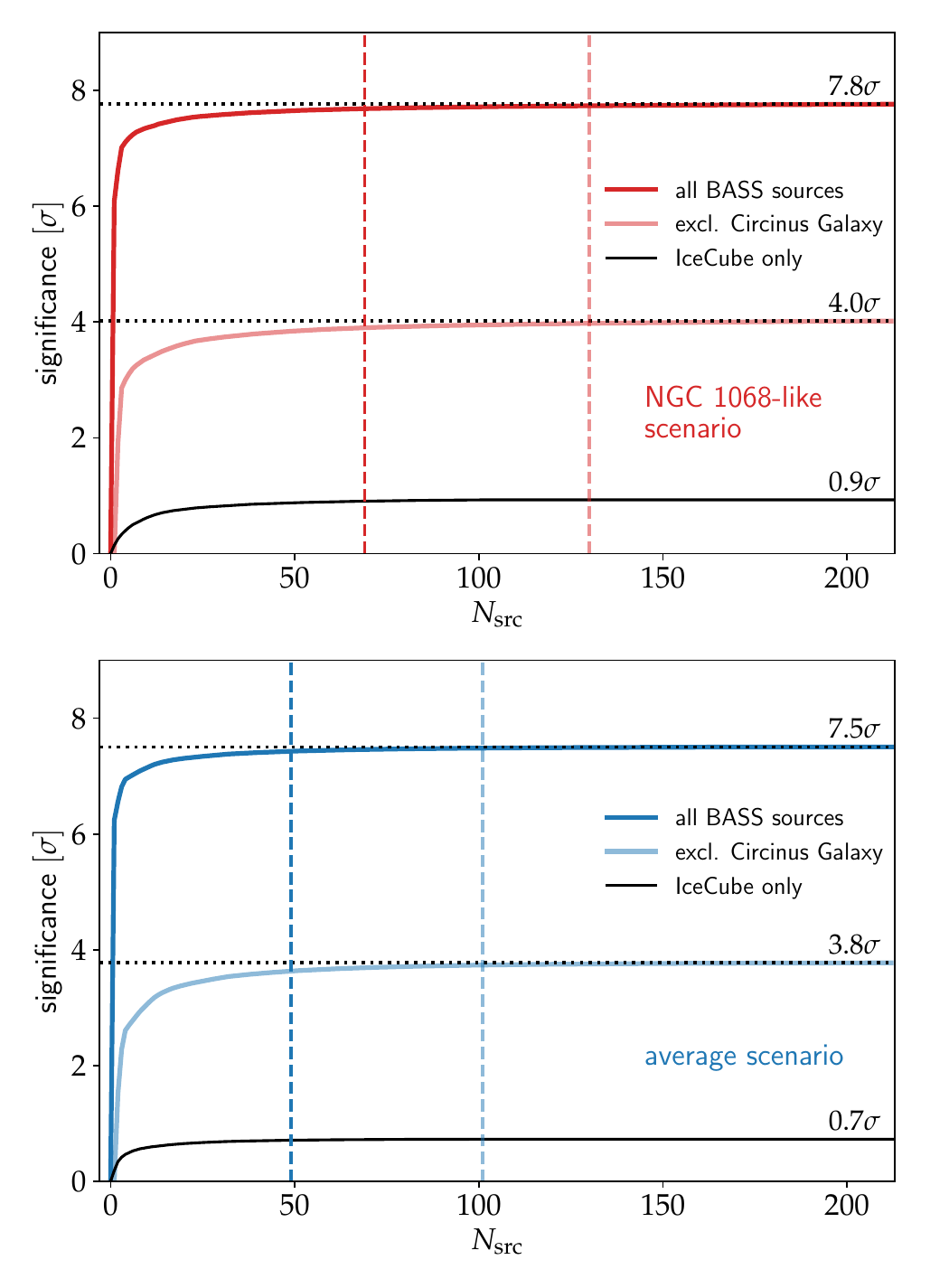}
\caption{Significance that would be achieved in a stacking analysis of all Seyfert galaxies in the BASS catalog with $F_X \geq 10^{-11}\,\mathrm{erg}\,\mathrm{cm}^{-2}\,\mathrm{s}^{-1}$ as a function of the number of stacked sources for the NGC~1068-like scenario (top panel) and the average scenario (bottom panel). Our predictions are made for 2040 and a detector network comprising IceCube, Baikal-GVD, KM3NeT/ARCA, and P-ONE. The results for the case in which the Circinus Galaxy is removed from the source sample are shown in light red and light blue, while the significance that would be obtained in a stacking analysis using IceCube data only is indicated by a black thin line. The vertical dashed lines mark the optimal number of stacked sources given in Table~\ref{tab:optimal_Nsrc}, and the black dotted lines correspond to the significance that is reached if all sources in the sample are included in the stacking analysis.}
\label{fig:sigma_vs_Nsrc}
\end{figure}

Figure~\ref{fig:p_val_vs_T} shows the $p$-value that would be achieved in the stacking analysis as a function of the observation time, taking into account the different start dates of those neutrino telescopes that are still under construction. While a stacking analysis using only IceCube data would not yield a significant result within a reasonable timeframe (see also Figure~\ref{fig:sigma_vs_Nsrc}), the sensitivity of the analysis increases drastically once Baikal-GVD and KM3NeT/ARCA, both located in the Northern hemisphere, start operating in their full configurations. This is mainly because 115 of the 213 sources considered in the analysis, including the four brightest Seyfert galaxies in the sample, are located in the Southern sky ($\delta < -5^\circ$) and are therefore not detectable by IceCube in the throughgoing-muon-track channel. In addition, both Baikal-GVD and KM3NeT/ARCA use water instead of ice as their detection medium and will thus have a better angular resolution than IceCube, especially at energies $> 10\,\mathrm{TeV}$ (see Figure~\ref{fig:ang_res_func} in Appendix~\ref{sec:ang_res_KM3NeT}).

As shown in Section~\ref{sec:point_src}, the Circinus Galaxy is predicted to be by far the brightest neutrino source in the sample and is expected to be detected as an individual point source by 2038. Therefore, we repeat the stacking analysis excluding Circinus. The results are summarized in the bottom two rows of Table~\ref{tab:results_stacking}. With only ${\sim}\hairsp 4\sigma$ in both scenarios, the significance that would be achieved in 2040 decreases drastically when the Circinus Galaxy is removed from the sample of stacked sources. Furthermore, a detection at the $5\sigma$ level would take ten years longer. This confirms that, if included, the Circinus Galaxy drives the result of the stacking analysis. Even in the case in which it is excluded, the start of Baikal-GVD and KM3NeT/ARCA leads to an increase in the sensitivity of the stacking analysis. However, this increase is slightly less pronounced, since we have removed the brightest source in the Southern sky from the sample. Nevertheless, we conclude that joint stacking analysis across multiple neutrino telescopes located at different geographical latitudes will be essential to go beyond the detection of a few, extraordinary Seyferts and to firmly establish Seyfert galaxies as a class of neutrino sources.

To determine the optimal stacking strategy, we investigate how the result of the stacking analysis depends on the number of stacked sources. To this end, we first sort all BASS catalog sources with $F_X \geq 10^{-11}\,\mathrm{erg}\,\mathrm{cm}^{-2}\,\mathrm{s}^{-1}$ according to their detectability as individual point sources. Then, we repeat the stacking analysis many times, gradually increasing the number of stacked sources starting from the source with the highest point-source significance. Figure~\ref{fig:sigma_vs_Nsrc} shows the significance that would be achieved in the stacking analysis as a function of the number of stacked sources, $N_\mathrm{src}$, for the NGC~1068-like scenario and the average scenario. In both scenarios, the significance stagnates after a certain number of sources have been included, since the additional, fainter sources contribute only negligible signal excess over the background. 

To quantify this effect, we determine the optimal number of sources to be stacked, which we define as the number of sources for which the significance is equal to 99\% of the significance that is achieved if all 213\footnote{212 sources if the Circinus Galaxy is excluded from the analysis.} sources in the sample are included in the analysis. The results of this calculation are summarized in Table~\ref{tab:optimal_Nsrc}. As indicated by the dashed vertical lines in Figure~\ref{fig:sigma_vs_Nsrc}, the optimal number of stacked sources is $N_\mathrm{src}\,{\sim}\, 50\,\text{--}\, 70$ if the Circinus Galaxy is included and $N_\mathrm{src}\,{\sim}\, 100\,\text{--}\, 130$ if the Circinus Galaxy is excluded, depending on the neutrino emission scenario under consideration. This suggests that stacking analyses including those ${\sim}\hairsp 100$ sources that are expected to be the brightest in neutrinos pose the optimal strategy to uncover the neutrino emission from Seyfert galaxies as a source class. Considering a much higher number of sources is not expected to lead to a more significant result.

The idea of a stacking analysis including a large number of faint neutrino sources, which is often referred to as \textit{deep stacking}, has also been studied by Kowalski \textit{et al.}~\cite{Kowalski2025}. Based on semi-analytical arguments, they demonstrate that in a scenario in which a few high-energy signal events dominate over a rather low background, deep stacking could lead to a significant increase of the detection sensitivity compared to conventional stacking analyses comprising a much lower number of sources. However, they also find that deep stacking is less promising in a high-background scenario, in which signal events at lower energies have to be identified over a dominating background. This is broadly consistent with our finding that stacking more than ${\sim}\hairsp 100$ sources does not lead to a significant improvement of the sensitivity of our analysis.

\begin{table}
    \caption{Optimal number of stacked sources including and excluding the Circinus Galaxy for both the NGC~1068-like scenario and the average scenario.}
    \begin{ruledtabular}
    \begin{tabular}{lcc}
        & incl.~Circinus & excl.~Circinus \\
        \hline
        \noalign{\vskip 1ex}
        NGC~1068-like scenario & 69 & 130 \\
        average scenario & 49 & 101 \\
    \end{tabular}
    \end{ruledtabular}
    \label{tab:optimal_Nsrc}
\end{table}

To visualize the Seyfert neutrino sky that would be seen by a detector network comprising IceCube, Baikal-GVD, KM3NeT/ARCA, and P-ONE, we use the \texttt{PLE$\nu$M} framework to calculate the expected number of neutrino events that would be detected by the network by 2040 from the entire population of Seyfert galaxies, including all BASS catalog sources considered in the stacking analysis. To this end, we simulate a population of Seyfert galaxies based on the X-ray luminosity function of AGNs by Ueda \textit{et al.}~\cite{Ueda2014} (see \citetalias{Saurenhaus2026} for more details), replace all simulated sources with $F_X \geq 10^{-11}\,\mathrm{erg}\,\mathrm{cm}^{-2}\,\mathrm{s}^{-1}$ by Seyfert galaxies from the BASS catalog, and compute the expected number of neutrino events from each source in the resulting population. In doing so, we take into account the anticipated start dates of the different neutrino telescopes listed in Table~\ref{tab:detector_props}. 

\begin{figure}
\includegraphics[width=8.5cm]{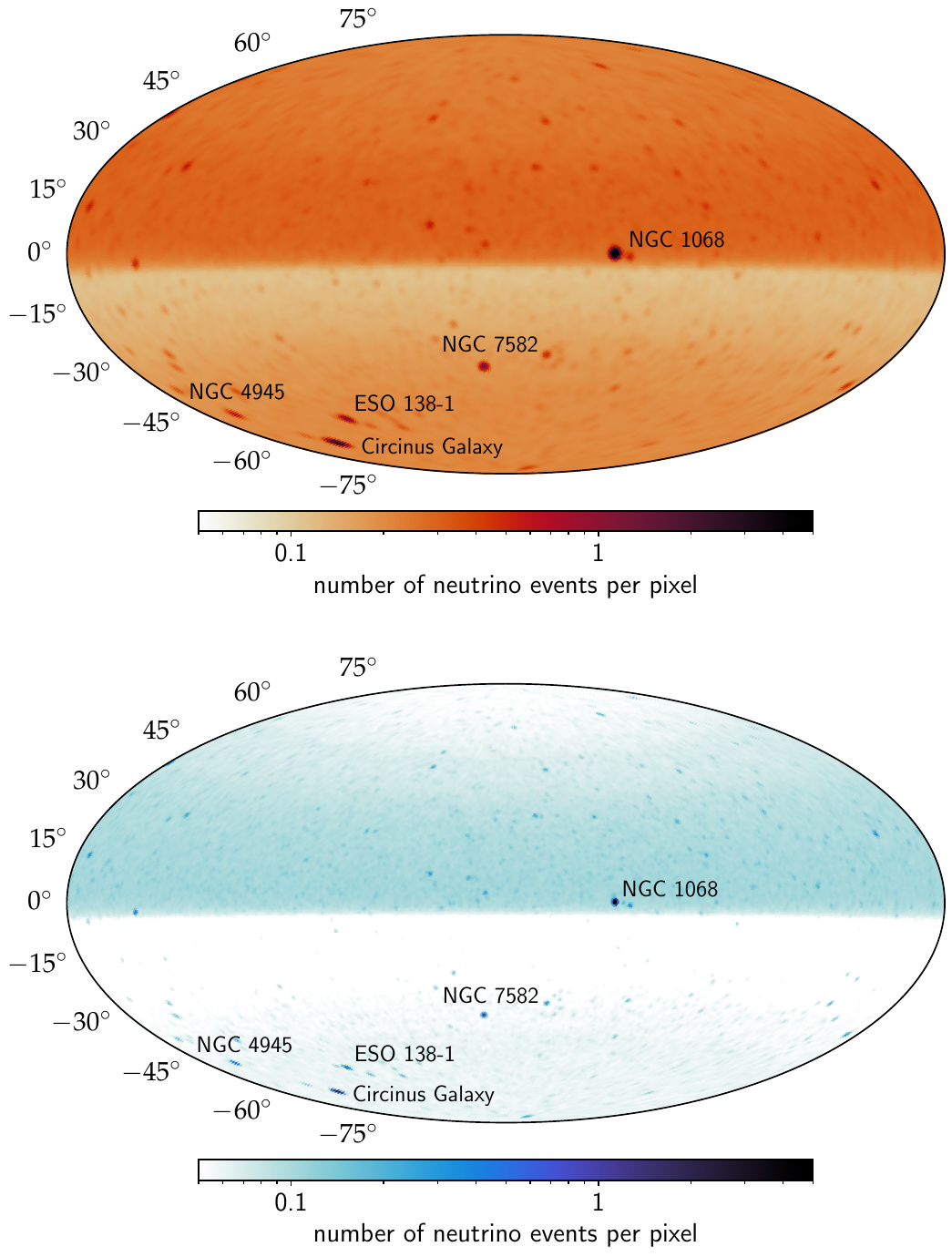}
\caption{Sky maps of the Seyfert neutrino sky expected to be seen by a detector network consisting of IceCube, Baikal-GVD, KM3NeT/ARCA, and P-ONE in 2040 in the NGC~1068-like scenario (top panel) and the average scenario (bottom panel), smoothed with a Gaussian kernel with $\sigma = 1^\circ$ and $\sigma = 0.5^\circ$, respectively. The color scale indicates the number of neutrino events per pixel, with darker colors corresponding to higher event numbers. In each panel, the five brightest sources are labelled. The sky maps show only neutrino events originating from the Seyfert population, and neither the atmospheric background nor possible contributions from other neutrino source classes are included.}
\label{fig:skymaps}
\end{figure}

Figure~\ref{fig:skymaps} shows the neutrino sky that the detector network would see in 2040 in the NGC~1086-like scenario and in the average scenario, excluding the background of atmospheric neutrinos and any potential contributions from other classes of neutrino sources. As shown in Figure~\ref{fig:nu_spectra_brightest_Seyferts}, in the NGC~1068-like scenario, the single-source neutrino spectra peak at ${\sim}\hairsp 1\,\mathrm{TeV}$, while in the average scenario, they peak at slightly higher energies of ${\sim}\hairsp 10\,\mathrm{TeV}$. To mimic the angular resolution of the neutrino telescopes in the network in the energy range in which the bulk of the neutrino events are detected, we smooth the sky maps with a Gaussian kernel with $\sigma = 1^\circ$ for the NGC~1068-like scenario and $\sigma = 0.5^\circ$ for the average scenario. These values roughly correspond to the angular resolution of KM3NeT/ARCA at $1\,\mathrm{TeV}$ and $10\,\mathrm{TeV}$, respectively, which does not differ significantly from that of IceCube at these energies \cite{Aiello2024,Abbasi2021}.

Overall, in the NGC~1068-like scenario considerably more neutrino events would be detected from the Seyfert population than in the average scenario. This is due to the fact that the underlying single-source neutrino spectra have a higher normalization. In addition, in both scenarios the number of detected events is highest in the Northern sky, while it is lowest for declinations $-40^\circ \lesssim \delta \lesssim -5^\circ$. Taking into account the different start dates of the considered neutrino telescopes, their cumulative exposure is much higher in the Northern sky than in the Southern sky because of the longer operation time of IceCube. Moreover, the effective area of Baikal-GVD, KM3NeT/ARCA, and P-ONE is slightly reduced for declinations $\delta \gtrsim -40^\circ$, since they can only observe this part of the sky in the throughgoing-muon-track channel for a fraction of the sidereal day. This results in a lower time-integrated exposure and therefore a lower expected number of detected neutrino events from declinations $-40^\circ \lesssim \delta \lesssim -5^\circ$. 

Nevertheless, small dark spots indicating positions from which an enhanced number of neutrino events is expected are visible in both the Northern and the Southern sky. Most of them correspond to nearby Seyfert galaxies from the BASS catalog, with the brightest ones being NGC~1068, the Circinus Galaxy, ESO~138-1, NGC~7582, and NGC~4945. Each of these sources produces $n_s \geq 15$ events in the NGC~1068-like scenario and $n_s \geq 5$ events in the average scenario. In addition, there are many fainter spots that are barely distinguishable from the background, which correspond to simulated sources that are slightly brighter than the population average. The diffuse emission from the remaining unresolved sources is visible as an orange background in the top panel and a light blue background in the bottom panel. 

% -------------------------------------------------------------------------------------------------

\section{Conclusions}
\label{sec:conclusions}

Based on the results of our population studies presented in \citetalias{Saurenhaus2026}, we have investigated the detectability of nearby, X-ray-bright Seyfert galaxies with present and future neutrino observatories for different neutrino emission scenarios. Besides NGC~1068, the Circinus Galaxy, located in the Southern sky, is expected to be by far the brightest Seyfert galaxy in both X-rays and neutrinos, making it one of the most interesting targets for the next-generation neutrino telescopes currently under construction in the Northern hemisphere. Considering a detector network comprising IceCube, Baikal-GVD, KM3NeT/ARCA, and P-ONE, we have shown that the Circinus Galaxy will be detected at the $5\sigma$ level by 2038 at the latest, provided that the Seyfert population is responsible for the diffuse astrophysical neutrino flux up to ${\sim}\hairsp 100\,\mathrm{TeV}$ and Circinus is at least as luminous in neutrinos as the average source in the population.

The majority of the remaining nearby Seyfert galaxies are predicted to be much fainter in neutrinos and are thus not expected to be detectable as individual point sources in the near future. Consequently, the most promising method to search for the neutrino emission from Seyfert galaxies as a source class will likely be a stacking analysis informed by X-ray observations. We have shown that in an average scenario, in which all sources are assumed to have a low pressure ratio of $P_\mathrm{CR}/P_\mathrm{th} = 0.01$ and highly turbulent coronae with $\eta = 1$, a stacking analysis of the brightest ${\sim}\hairsp 50$ Seyfert galaxies combining data from IceCube, Baikal-GVD, KM3NeT/ARCA, and P-ONE would result in a $5\sigma$ discovery already in 2036, only 6 years after the anticipated start date of KM3NeT/ARCA. In contrast, a stacking analysis using IceCube data only is not expected to yield a significant result within a reasonable timeframe. These results show that next-generation neutrino telescopes will play a key role in advancing our understanding of Seyfert galaxies as potential neutrino sources and demonstrate the importance of collaboration efforts across different experiments for a successful and efficient search for the sources of the astrophysical neutrino flux.

Additionally, we have shown that $\gamma$-ray observations in the MeV range provide a means for testing the assumption that the neutrinos associated with NGC~1068 and other nearby Seyfert galaxies originate from the AGN corona. The $\gamma$-ray telescope COSI \cite{Tomsick2023}, which is scheduled for launch already in 2027, will be sensitive enough to test our model predictions for NGC~1068, and possibly also for some of the brightest sources in the Southern sky, within an observation time of just a few years. AMEGO-X \cite{Caputo2022} and newASTROGAM \cite{Berge2025} are planned to have a significantly higher sensitivity and would thus enable observations of a larger number of Seyfert galaxies, which are crucial for improving our understanding of Seyfert galaxies as hidden neutrino sources.

% -------------------------------------------------------------------------------------------------

\medskip
\noindent
\textit{Note added.} While we were finalizing this work, we became aware of a related study by Mukhopadhyay \textit{et al.}~\cite{Mukhopadhyay2026}. The two works have been carried out independently and are complementary. Mukhopadhyay \textit{et al.} focus on building a sample of AGN neutrino candidates based on X-ray observations in different energy bands and subsequently evaluate the detectability of the sources in their sample with IceCube-Gen2 and KM3NeT. In contrast, we consider a network of neutrino telescopes comprising IceCube, Baikal-GVD, KM3NeT/ARCA, and P-ONE, and model the instrument response of the individual experiments, including their energy and angular resolution. In addition, we take into account constraints from $\gamma$-ray observations in the MeV range and explore the potential of deep stacking of a large number of X-ray-bright Seyfert galaxies.

% -------------------------------------------------------------------------------------------------

\appendix

\section{Angular resolution function of KM3NeT/ARCA}
\label{sec:ang_res_KM3NeT}

The angular resolution of a neutrino telescope provides a measure for how close a muon is reconstructed to the arrival direction of its parent neutrino. It is a function of the true neutrino energy, $E_\nu$, and the angular distance between the reconstructed muon direction, $\hat{\Omega}$, and the true arrival direction of the neutrino, $\Omega_\nu$, which is defined as $\Psi = |\hat{\Omega} - \Omega_\nu|$. In this work, we model the angular resolution function of KM3NeT/ARCA based on the angular deviation reported in Ref.~\cite{Aiello2024}, which has been obtained from simulations of the full KM3NeT/ARCA detector.

\begin{figure}
    \includegraphics[width=7.5cm]{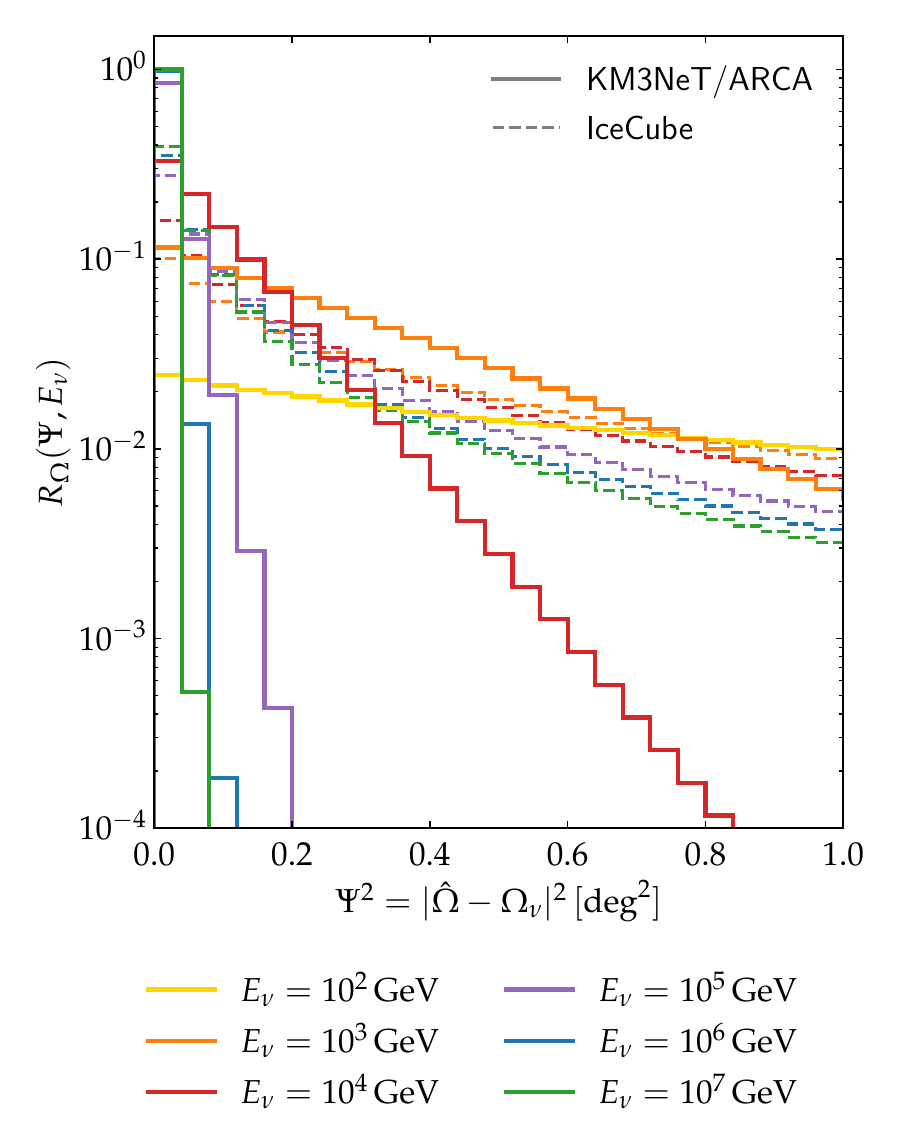}
    \caption{Angular resolution function of KM3NeT/ARCA (solid lines) and IceCube (dashed lines) binned in true neutrino energy and the squared angular distance between the reconstructed muon direction and the true arrival direction of the neutrino. The angular resolution function of KM3NeT/ARCA is computed from the angular deviation reported in Ref.~\cite{Aiello2024} using Eqs.~(\ref{eq:Rayleigh_func}) and (\ref{eq:ang_dev}), while that of IceCube is taken from the public 10-year data release \cite{IceCube2021} as implemented in \texttt{PLE$\nu$M}.}
    \label{fig:ang_res_func}
\end{figure}

Based on the assumption that the distribution of the reconstructed directions of secondary muons follows a 2D Gaussian with respect to the true arrival direction of the parent neutrino, the angular resolution function for a fixed neutrino energy can be modeled as a Rayleigh distribution of the form 
\begin{equation}
    R_\Omega (\Psi, E_\nu) = \frac{\Psi}{\sigma^2} e^{-\Psi^2/2\sigma^2}
    \label{eq:Rayleigh_func}
\end{equation}
with the scale parameter $\sigma \equiv \sigma (E_\nu)$ (see e.g.~\cite{Glauch2021}). The angular deviation, $\tilde{\psi}$, is equivalent to the median opening angle between the reconstructed muon direction and the true neutrino direction and related to the scale parameter as
\begin{equation}
    \tilde{\psi} = \sigma \sqrt{2\log 2}\,.
    \label{eq:ang_dev}
\end{equation}
The resulting angular resolution function, binned in both $\Psi^2 \equiv |\hat{\Omega}-\Omega_\nu|^2$ and $E_\nu$, is shown in Figure~\ref{fig:ang_res_func}. Since Ref.~\cite{Aiello2024} only reports the angular deviation for KM3NeT/ARCA for energies $E_\nu \,{\geq}\, 1\,\mathrm{TeV}$, we conservatively approximate its angular resolution function at lower energies by that of IceCube. For neutrino energies below ${\sim}\hairsp 10\,\mathrm{TeV}$, the angular resolution of any neutrino telescope is anyway limited by the non-negligible kinematic opening angle between the incident neutrino and the secondary muon, independent of the detection medium \cite{Learned2000}.

Overall, a narrower angular resolution function results in a neutrino being reconstructed closer to the position of its source in the sky. This leads to a higher signal-to-background ratio in the direction of the source and thus to a higher discovery potential. For higher neutrino energies, the angular resolution functions of both IceCube and KM3NeT/ARCA are more peaked towards $\Psi^2 = 0$, reflecting the fact that secondary muons produced by high-energy neutrinos are reconstructed closer to the true arrival direction of the neutrino than those produced by low-energy neutrinos. Although KM3NeT/ARCA is expected to have a better angular resolution than IceCube due to the fact that it uses water instead of ice as its detection medium, for energies $1\,\mathrm{TeV} \lesssim E_\nu \lesssim 10\,\mathrm{TeV}$, the angular resolution of KM3NeT/ARCA appears to be slightly worse than that of IceCube (see orange solid and dashed lines in Figure~\ref{fig:ang_res_func}). This discrepancy likely arises because we construct the angular resolution function of KM3NeT/ARCA using only the median angular deviation, making the simplifying assumption that for a fixed neutrino energy, it can be approximated by a Rayleigh distribution. In contrast, the angular resolution function of IceCube implemented in \texttt{PLE$\nu$M} is based on detector simulations of IceCube and built using Gaussian kernel density estimation, which is more accurate.

% -------------------------------------------------------------------------------------------------

\section{Neutrino and MeV $\gamma$-ray spectra for the remaining sources}
\label{sec:nu_and_gamma_spec_remaining_srcs}

\begin{figure*}[p]
    \includegraphics[width=17.5cm]{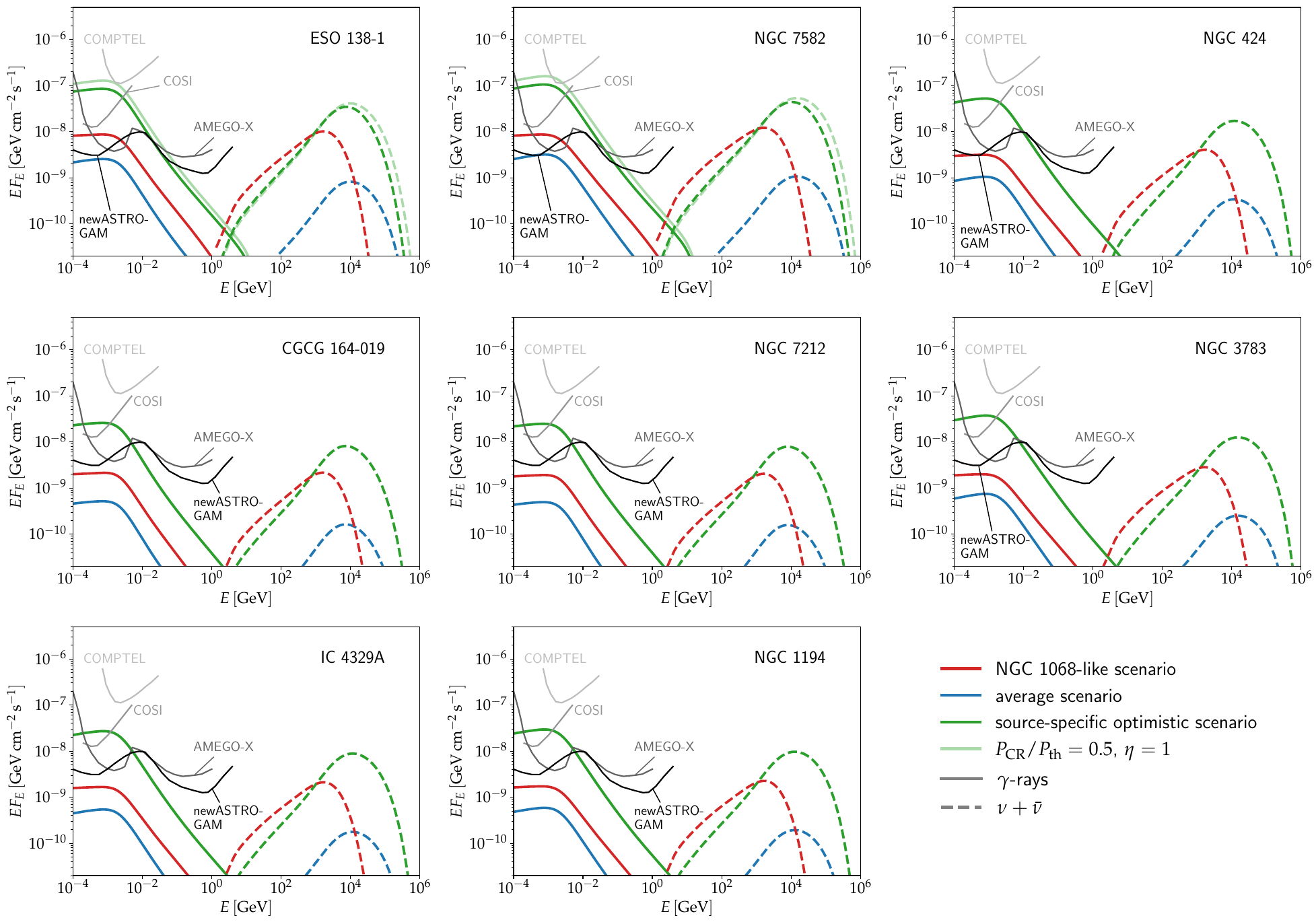}
    \caption{All-flavor neutrino (dashed lines) and $\gamma$-ray spectra (solid lines) for the NGC~1068-like scenario (red), the average scenario (blue), and the most optimistic scenario that is compatible with the non-detection of the respective source by COMPTEL (green) for ESO~138-1, NGC~7582, NGC~424, CGCG~164-019, NGC~7212, NGC~3783, IC~4329A, and NGC~1194. For those sources for which the non-detection by COMPTEL constrains the allowed values of $P_\mathrm{CR}/P_\mathrm{th}$ and $\eta$, the neutrino and $\gamma$-ray spectra for $P_\mathrm{CR}/P_\mathrm{th} = 0.5$ and $\eta = 1$ are shown separately in light green. In addition, the sensitivity of COMPTEL \cite{Takahashi2013} is shown for its total lifetime of ${\sim}\hairsp 9\,\mathrm{years}$, while the sensitivity curves of COSI \cite{Tomsick2023}, AMEGO-X \cite{Caputo2022}, and newASTROGAM \cite{Berge2025} are shown for an observation time of $3\,\mathrm{years}$.}
    \label{fig:spectra_8_sources}
\end{figure*}

Figure~\ref{fig:spectra_8_sources} shows the neutrino and MeV $\gamma$-ray spectra for those eight Seyfert galaxies in Table~\ref{tab:src_properties} that have not been detected in $\gamma$-rays so far. In the population-wide optimistic scenario ($\eta = 1$ and $P_\mathrm{CR}/P_\mathrm{th} = 0.5$), the cascaded $\gamma$-ray emission from ESO~138-1 and NGC~7582 would exceed the integrated sensitivity of COMPTEL at ${\sim}\hairsp 1\,\mathrm{MeV}$. Yet, neither of these two sources was in fact detected by COMPTEL. This allows to constrain the parameter values describing their source-specific optimistic scenario to $\eta = 2$ and $P_\mathrm{CR}/P_\mathrm{th} = 0.5$ for each source. For all other sources, our $\gamma$-ray predictions for $\eta = 1$ and $P_\mathrm{CR}/P_\mathrm{th} = 0.5$ are consistent with a non-detection by COMPTEL and require no further adjustment.

% -------------------------------------------------------------------------------------------------

\section{Detailed results for individual sources}
\label{sec:detailed_results}

In Section~\ref{sec:results_point_sources}, we present our detectability estimates for the ten Seyfert galaxies listed in Table~\ref{tab:src_properties} as individual point sources for a neutrino telescope network consisting of IceCube, Baikal-GVD, KM3NeT/ARCA, and P-ONE in the year 2040. Tables~\ref{tab:ps_detectability_NGC_like_and_avg_scen} and \ref{tab:ps_detectability_opt_scen} provide a complete summary of our results for the NGC~1068-like scenario, the average scenario, and the source-specific optimistic scenario. Listed are the expected number of signal events, the TS values, the $p$-values together with the equivalent significance, the years in which a significance of $3\sigma$ and $5\sigma$ would be reached, and, for the source-specific scenario, the most optimistic values of $P_\mathrm{CR}/P_\mathrm{th}$ and $\eta$ for which the cascaded $\gamma$-ray emission predicted by our model does not violate any $\gamma$-ray observations of the respective source. 

\onecolumngrid

\begin{table*}[p]
    \caption{Detectability of the ten brightest Seyfert galaxies in the BASS catalog (excl.~NGC~1068) with a detector network consisting of IceCube, Baikal-GVD, KM3NeT/ARCA, and P-ONE in 2040 for the NGC~1068-like scenario and the average scenario. For each source and scenario, we list the expected number of signal events, the TS value, the $p$-value together with the equivalent significance, and the years in which a significance of $3\sigma$ and $5\sigma$ would be reached.}
    \begin{ruledtabular}
    \begin{tabular}{lrrrrrrrrrr}
        & \multicolumn{5}{c}{NGC~1068-like scenario} & \multicolumn{5}{c}{average scenario} \\
        \cline{2-6} \cline{7-11}
        \noalign{\vskip 0.75ex}
        & \multicolumn{1}{c}{$n_s$} & \multicolumn{1}{c}{TS} & \multicolumn{1}{c}{$-\log_{10} p$} & {$3\sigma$ in} & {$5\sigma$ in} & \multicolumn{1}{c}{$n_s$} & \multicolumn{1}{c}{TS} & \multicolumn{1}{c}{$-\log_{10} p$} & {$3\sigma$ in} & {$5\sigma$ in} \\
        \hline
        \noalign{\vskip 0.75ex}
        Circinus Galaxy & 98.3 & 41.1 & 8.9 ($6.1\sigma$) & 2033 & 2038 & 26.0 & 43.1 & 9.3 ($6.2\sigma$) & 2033 & 2038\\
        ESO~138-1 & 30.1 & 5.2 & 1.1 ($1.8\sigma$) & 2052 & ${>}\hairsp 2061$ & 5.4 & 1.8 & 0.4 ($0.8\sigma$) & ${>}\hairsp 2061$ & ${>}\hairsp 2061$ \\
        NGC~7582 & 33.1 & 5.7 & 1.2 ($1.9\sigma$) & 2050 & ${>}\hairsp 2061$ & 7.0 & 3.5 & 0.8 ($1.4\sigma$) & 2061 & ${>}\hairsp 2061$ \\
        NGC~424 & 10.6 & 0.6 & 0.1 ($0.3\sigma$) & ${>}\hairsp 2061$ & ${>}\hairsp 2061$ & 2.1 & 0.4 & 0.1 ($0.2\sigma$) & ${>}\hairsp 2061$ & ${>}\hairsp 2061$ \\
        CGCG~164-019 & 8.9 & 0.2 & 0.0 ($0.0\sigma$) & ${>}\hairsp 2061$ & ${>}\hairsp 2061$ & 1.6 & 0.0 & 0.0 ($0.0\sigma$) & ${>}\hairsp 2061$ & ${>}\hairsp 2061$ \\
        NGC~7212 & 8.8 & 0.1 & 0.0 ($0.0\sigma$) & ${>}\hairsp 2061$ & ${>}\hairsp 2061$ & 1.7 & 0.0 & 0.0 ($0.0\sigma$) & ${>}\hairsp 2061$ & ${>}\hairsp 2061$ \\ 
        NGC~4945 & 16.1 & 0.8 & 0.2 ($0.4\sigma$) & ${>}\hairsp 2061$ & ${>}\hairsp 2061$ & 5.6 & 4.1 & 0.9 ($1.5\sigma$) & 2057 & ${>}\hairsp 2061$ \\ 
        NGC~3783 & 7.3 & 0.3 & 0.1 ($0.2\sigma$) & ${>}\hairsp 2061$ & ${>}\hairsp 2061$ & 1.6 & 0.3 & 0.1 ($0.2\sigma$) & ${>}\hairsp 2061$ & ${>}\hairsp 2061$ \\ 
        IC~4329A & 4.8 & 0.1 & 0.0 ($0.0\sigma$) & ${>}\hairsp 2061$ & ${>}\hairsp 2061$ & 0.9 & 0.1 & 0.0 ($0.0\sigma$) & ${>}\hairsp 2061$ & ${>}\hairsp 2061$ \\ 
        NGC~1194 & 9.1 & 0.2 & 0.0 ($0.0\sigma$) & ${>}\hairsp 2061$ & ${>}\hairsp 2061$ & 2.3 & 0.1 & 0.0 ($0.0\sigma$) & ${>}\hairsp 2061$ & ${>}\hairsp 2061$ \\
    \end{tabular}
    \end{ruledtabular}
    \label{tab:ps_detectability_NGC_like_and_avg_scen}
\end{table*}

\begin{table*}[h]
    \caption{Detectability of the ten brightest Seyfert galaxies in the BASS catalog (excl.~NGC~1068) with a detector network consisting of IceCube, Baikal-GVD, KM3NeT/ARCA, and P-ONE in 2040 for the source-specific optimistic scenario compatible with the $\gamma$-ray observations of the respective source. For each source, we list the corresponding values of the model parameters $\eta$ and $P_\mathrm{CR}/P_\mathrm{th}$ as well as the expected number of signal events, the TS value, the $p$-value together with the equivalent significance, and the years in which a significance of $3\sigma$ and $5\sigma$ would be reached.}
    \begin{ruledtabular}
    \begin{tabular}{lrcrrrcc}
        & \multicolumn{1}{c}{$\eta$} & {$P_\mathrm{CR}/P_\mathrm{th}$} & \multicolumn{1}{c}{$n_s$} & \multicolumn{1}{c}{TS} & \multicolumn{1}{c}{$-\log_{10} p$} & {$3\sigma$ in} & {$5\sigma$ in} \\
        \hline
        \noalign{\vskip 1ex}
        Circinus Galaxy & 15 & 0.5 & 432.6 & 1099.3 & 238.7 ($33.0\sigma$) & 2029 & 2029 \\
        ESO~138-1 & 2 & 0.5 & 205.8 & 452.4 & 98.2 ($21.1\sigma$) & 2029 & 2030 \\
        NGC~7582 & 2 & 0.5 & 261.4 & 700.8 & 152.2 ($26.3\sigma$) & 2029 & 2029 \\
        NGC~424 & 1 & 0.5 & 105.6 & 210.4 & 45.7 ($14.3\sigma$) & 2030 & 2031 \\
        CGCG~164-019 & 1 & 0.5 & 77.5 & 45.5 & 9.9 ($6.4\sigma$) & 2025 & 2035 \\ 
        NGC~7212 & 1 & 0.5 & 84.3 & 47.6 & 10.3 ($6.6\sigma$) & 2029 & 2035 \\
        NGC~4945 & 10 & 0.5 & 132.7 & 226.8 & 49.2 ($14.9\sigma$) & 2030 & 2031 \\
        NGC~3783 & 1 & 0.5 & 79.8 & 154.4 & 33.5 ($12.2\sigma$) & 2030 & 2032 \\
        IC~4329A & 1 & 0.5 & 47.2 & 65.4 & 14.2 ($7.8\sigma$) & 2032 & 2035 \\
        NGC~1194 & 1 & 0.5 & 112.7 & 101.1 & 21.9 ($9.8\sigma$) & 2021 & 2030 \\
    \end{tabular}
    \end{ruledtabular}
    \label{tab:ps_detectability_opt_scen}
\end{table*}

\twocolumngrid

% -------------------------------------------------------------------------------------------------

\bibliography{references.bib}

\end{document}